\documentclass[a4paper,11pt]{article}

\usepackage{jcappub}
\usepackage[english]{babel}
\usepackage[letterpaper,top=2cm,bottom=2cm,left=3cm,right=3cm,marginparwidth=1.75cm]{geometry}
\usepackage{mathtools}
\usepackage{nicefrac}
\usepackage{amsmath,amssymb}
\usepackage{xcolor}
\usepackage{graphicx}
\usepackage{subcaption}
\usepackage{float}
\usepackage{xfrac}

\newcommand{\dd}{\mathrm{d}}
\newcommand{\N}{{\mathchoice{}{}{\scriptscriptstyle}{}N}}

\def\ps{\mathcal{P}_{_{\mathrm{R}}}}
\def\pt{\mathcal{P}_{_{\mathrm{T}}}}

\title{\boldmath Spectral Distortion Signatures of Step-like Inflationary Potential}

\author[a,b]{Jorge Mastache}
\author[c]{Wilson Barrera}
\author[c]{Raúl Henríquez-Ortiz}

\affiliation[a]{Consejo Nacional de Humanidades, Ciencia y Tecnolog\'ia, Av. Insurgentes Sur 1582, Colonia Cr\'edito Constructor, Del. Benito Ju\'arez, 03940, Ciudad de M\'exico, M\'exico} 
\affiliation[b]{Mesoamerican Centre for Theoretical Physics, Universidad Aut\'{o}noma de Chiapas,  Carretera Zapata Km. 4, Real del Bosque, 29040, Tuxtla Guti\'{e}rrez, Chiapas, M\'{e}xico,}
\affiliation[c]{Escuela de F\'isica, Facultad de Ciencias Naturales y Matem\'atica, Universidad de El Salvador,\\ final de Av.\ M\'artires y H\'eroes del 30 julio, San Salvador, CP 1101, El Salvador.}

\emailAdd{jh.mastache@mctp.mx}
\emailAdd{bv17003@ues.edu.sv}
\emailAdd{raul.henriquez@ues.edu.sv}

\abstract{In this work, we analyze a power-law inflationary potential enhanced with a step that can introduce features in the primordial power spectrum. We focus on the computation of the Spectral Distortions (SD) induced by these features obtained from the inflationary dynamics. In this scenario, we explore the potential of upcoming experimental missions like PIXIE to detect the SD of the model within a power of $n = \nicefrac{2}{3}$, a power that agrees with recent tensor-to-scalar ratio constraints. The model offers insights into models with cosmological phases and different scalar field dynamics. Introducing a step in the inflaton potential leads to distinct features in the primordial power spectrum, such as oscillations and localized enhancements/suppressions at specific scales. We analyze the impact of three primary parameters—$\beta$, $\delta$, and $\phi_{\text{step}}$—on the amplitude and characteristics of the SD. The $\phi_{\rm step}$ places the onset of the oscillations in the primordial power spectrum. The $\beta$ parameter significantly influences the magnitude of the $\mu$-SD, with its increase leading to larger SD and vice versa. Similarly, the $\delta$ parameter affects the smoothness of the step in the potential, with larger values resulting in smaller SD. Our findings indicate a distinct parameter space defined by $0.02 <\delta/{\rm M_{pl}} \lesssim 0.026$, $0.10 \lesssim \beta < 0.23$, and $ 7.53 \lesssim \phi_{\rm step}/{\rm M_{pl}} \lesssim 7.55$, which produces SD potentially detectable by PIXIE. This region also corresponds to the maximum observed values of $\mu$ and $y$ SD, which in special cases are an order of magnitude larger than the expected for $\Lambda$CDM. However, we also identify parameter ranges where $\mu$ and $y$ SD may not be detectable due to the limitations of current observational technology. This comprehensive analysis of SD provides constraints of step-like inflationary models and their implications on its dynamics.   }

\begin{document}
\maketitle
\flushbottom

\section{Introduction}  

Cosmological inflation is a proposed mechanism to explain the observed properties of our Universe that remain challenging within the standard Big Bang model. Specifically, it addresses the flatness problem that refers to the curvature of the Universe on large scales, and the horizon problem, which notes the fact that different regions of the Universe that now are not in casual contact appear to have the same observational properties \cite{Guth:1980zm, Linde:1981mu, linde1983}. Inflation proposes a brief period of rapid, exponential expansion shortly after the Big Bang. This theory predicts the observed isotropy and homogeneity of the Universe as observed in the cosmic microwave background radiation (CMB) \cite{Planck:2018vyg, Planck:2018jri}. Furthermore, inflation accounts for the primordial density fluctuations that seeded the formation of galaxies and cosmic structures.

The inflationary model is driven by a scalar field known as the inflaton, a homogeneous scalar field $\phi$, with an associated potential $V(\phi)$. Quantum fluctuations in this inflaton field, $\delta\phi$, expand on cosmological scales and establish the initial condition for the cosmological perturbation theory. These fluctuations, combined with the predicted isotropy and homogeneity, can be empirically verified through CMB observations, large-scale structure surveys, and the primordial power spectrum \cite{Planck:2018jri}.

Spectral distortions refer to deviations from the near-perfect blackbody spectrum of the CMB \cite{Chluba:2019kpb}. Two primary types of CMB spectral distortions dominate: (1) High-energy electrons interact with CMB photons via inverse Compton scattering, leading to photon thermalization. As the Universe cools down, the diminishing efficiency of the Compton scattering results in a shift of the spectra towards lower frequencies, known as $\mu$-type SD \cite{Zeldovich:1969ff, Illarionov1975S}. (2) When CMB photons scatter off high-energy electrons in a hot, ionized gas cloud the photons gain energy from an inverse Compton scattering process resulting in the $y$-type SD \cite{Bianchini:2022dqh, Ravenni:2017lgw, Chluba:2011hw}. There is a third kind of distortion, usually called the $r$-type, which depicts the transition between the $\mu$ and $y$ SD \cite{Khatri:2012tw}. 

These distortions can provide important information about the properties of the distribution of matter in the Universe and, therefore, are subject to the initial conditions of the clustering of the matter. The energy injection, thermalization, and recombination of primordial particles are the standard phenomena in the early Universe that can produce SD \cite{Lucca:2019rxf, Ali-Haimoud:2021lka, Chluba:2013dna, Chluba:2015hma, Khatri:2013dha}. Silk damping, a phenomenon resulting from the diffusion of photons and baryons, leads to the dissipation of energy storage in the acoustic waves. This process offers an avenue to investigate the primordial power spectrum, thereby offering a way to study inflationary dynamics \cite{Chluba:2012gq, Chluba:2016bvg, Cabass:2018jgj, Chluba:2019kpb, Clesse:2014pna}.  Spectral distortions offer a complementary avenue at small scales, $k \gtrsim 1 \, \mathrm{Mpc^{-1}}$, which are not accessible through CMB anisotropies and polarization, allowing us to access distinct cosmological information and potentially constraining the nature of the inflationary period within the detection capabilities of forthcoming experiments that expect to put constraints to the primordial power spectrum (PPS) at scales of $1 \lesssim k/\mathrm{Mpc^{-1}} \lesssim \mathcal{O}(10^4)$ \cite{Khatri:2013dha, Chluba:2012we}. 

The canonical inflationary model predicts a gaussian, smooth, and featureless PPS that can be described with a power--law parametrization, and current reconstructions for the PPS using CMB data and large-scale structure data statistically agree to these predictions \cite{Wang:2022nml, Sohn:2022jsm, Planck:2018jri, Palma2018, Hunt2015, Hazra2014, Hu2014, Aslanyan2014, Hunt2014, Hazra2013b, Hazra2013, Paykari2014, Aslanyan2014, Hunt2014, Vazquez2012, Gauthier2012, Hamann2010, Peiris:2009wp, Ichiki2009, Park:2008hz, Bridges:2008ta, Verde2008, Leach2006, Sealfon:2005em, TocchiniValentini2005, Shafieloo2004, Mukherjee:2003cz, Hannestad:2003zs, Hazra:2010ve}. However,  departures from a scale-invariant PPS have been observed, but they should be approached with caution because it remains challenging to distinguish if the feature underpins inflationary models or if it is a statistical artifact \cite{Palma2018, Hazra2014, Hunt2015, Hu2014, Aslanyan2014, Hunt2014, Verde2008, Leach2006, TocchiniValentini2005, Mukherjee:2003cz, Lodha:2023jru, Hazra:2010ve}. As cosmological data continues to refine, the PPS offers observational constraints on inflationary models, richer structures in the inflaton potential introduce features in the PPS, such as oscillations or localized enhancements/suppressions at certain scales. The nature of the inflaton field determines the scale, amplitude, and form of these features that can even be located outside of the observable window. 
 
The specific nature of the mechanism driving inflation is still unknown; hence, we aim to explore more complex inflationary dynamics using the SD. In particular, we aim to forecast the feasibility of the step-potential inflationary models \cite{Adams:2001vc, Ichiki:2009xs, Chen:2008wn, Kusenko:1997si} using SD. Incorporating a step in the inflaton potential allows for the exploration of specific particle physics scenarios and their implications for the early Universe. For instance, it can simulate a phase transition resulting from a change in a system symmetry \cite{Kusenko:1997si}; it can depict non-trivial quantum field theory effects that may be relevant during inflation \cite{Chen:2008wn}; the step can arise from the presence of additional scalar fields associated with supersymmetric partners, leading to testable predictions \cite{Allahverdi:2010xz}. Within string theory, particular compactifications or brane configurations can induce steps or kinks from geometric or topological properties of the compactified dimensions \cite{Baumann:2014nda}. Moreover, introducing a step in the inflaton potential can modify the dynamics, potentially avoiding the trans-Planckian problem and offering a consistent framework to describe the behavior of the inflaton field \cite{Martin:2000xs, Brandenberger:2012aj}. Thus, examining the step potential can offer a comprehensive understanding of the underlying particle physics in the early Universe.
 
Both the $\mu$-type and $y$-type spectral distortions are very subtle, with amplitudes on the order of $\mu = \mathcal{O}(10^{-8})$ and $y =\mathcal{O}(10^{-9})$, respectively. However, future experiments such as the Primordial Inflation Explorer (PIXIE) mission \cite{Khatri:2013dha, Kogut:2011xw, kogut2016}, the Cosmic Origins Explorer (COrE) satellite \cite{COrE:2011bfs}, LiteBIRD \cite{LiteBIRD:2022cnt, Matsumura:2013aja}, PICO \cite{NASAPICO:2019thw}, the Voyage 2050 \cite{Chluba:2019nxa}, CMB-S4 \cite{CMB-S4} or PRISM \cite{PRISM:2013ybg, PRISM:2013fvg}; which are expected to improve the sensitivity by several orders of magnitude to measure both SD. The SD will provide a unique window into the physics of the early Universe and inflation. Therefore, it is important to forecast interesting scenarios for this observation.

In this paper, we aim to provide constraints for the inflaton with a chaotic potential with a step and forecast observational evidence of the spectral distortions in this scenario. We will delve into the dynamics of inflation and the introduction of the step-potential in Sec.\ref{sec:inflation_dynamic}, then, in Sec.\ref{sec:spectral_distortions} an overview of the theory of SD for primordial small-scale perturbations is presented.  With all these previous elements, in Sec.\ref{sec:results}, we compute and discuss the observable features of SD for the studied scenarios. Finally, we summarize our main results in Sec.\ref{sec:conclutions}.

\section{Background Equation of Motion} \label{sec:inflation_dynamic}   

We assume that inflation is driven by a scalar field called inflaton, $\phi(t)$, on a flat, isotropic, and homogeneous background described by the Friedmann-Lema\^{i}tre-Robertson-Walker (FLRW) metric: 
\begin{equation}
    \dd s^{2}= -\dd t^{2}+a(t)^{2}\delta_{ij}\dd x^{i}\dd x^{j}\,,
    \label{eq:metric_frw}
\end{equation}
where $a$ is the scale factor. The dynamics of the Universe driven by the inflaton field is given by the Friedmann and Klein-Gordon (KG) equations
\begin{eqnarray} 
    && H^{2} = \frac{\rho_{\phi}}{3M_{pl}^{2}} \,,\label{eq:friedmann}\\
    && \ddot{\phi}+3H\dot{\phi}+V^\prime_{\phi} = 0 \; , \label{eq:KG}
\end{eqnarray}
where dots are derivatives with respect to time while primes are the derivatives with respect to the field, $V^\prime_{\phi} = dV(\phi)/d\phi$; $H=\dot{a}/a$ is the Hubble parameter, $M_{pl} = 1/\sqrt{8\pi G}$ is the reduced Planck mass. From the energy momentum-tensor, the energy density $\rho_\phi$ and pressure $P_\phi$ on the inflaton is given by $\rho_{\phi}=\dot{\phi}^{2}/2+V(\phi)$ and $P_\phi=\dot{\phi}^{2}/2-V(\phi)$. 

The number of e-fold, $N$, quantifies the amount of exponential growth that occurs during inflation, defined as the logarithm of the ratio of the scale factor at the end of inflation ($a(t_e)$) and at a time $t$, $a(t)$,
\begin{equation}
    N \equiv \ln \frac{a(t_{e})}{a(t)} =\int_{t}^{t_{e}}H \dd t \; . 
    \label{eq:e_folds}
\end{equation}
The minimum number of e-folds that solve the horizon and flatness problems is $N \approx 60$. Writing the equation using $N$ as an evolution variable allows the optimization of the computational analysis; therefore, the evolution of the background equations will be written as a function of e-folds, making use of the relation $dN = H dt$ \cite{Liu:2010dh, Liu:2012iba}. The subscript $N$ on a function will depict the derivative with respect to the number of e-folds, e.g., for any function $f$ we have $f_N = df/dN$.

Slow-roll is a requirement for inflation that ensures that the expansion is sustained for a sufficient time; while slow-rolling, the kinetic energy of the inflaton remains small compared to its potential energy. One parameter that characterizes slow-roll is $\epsilon$ defined as
\begin{equation}
    \epsilon = -\frac{\dot{H}}{H^2}=-\frac{H_N}{H}
    \label{eq:epsilon}
\end{equation}
Finally, inflation lasts while $\epsilon < 1$.

The KG equation, Eq.\eqref{eq:KG}, which describes the dynamics of the inflaton field, is written in terms of the $N$ as 
\begin{equation}
    \phi_{N N}+\left(\frac{H_N}{H}+3\right) \phi_N+\frac{1}{H^2} \frac{\partial V(\phi)}{\partial \phi}=0
    \label{eq:KG_f_N}
\end{equation}
Combining the Friedmann, Eq.\eqref{eq:friedmann}, with the continuity equations, $\dot{\rho}_\phi = -3H(\rho_\phi + P_\phi)$, we obtained the equation
\begin{eqnarray}
    \dot{H}=-\frac{\dot{\phi}^2}{2 M_{\mathrm{pl}}^2} \quad \textrm{or} \quad  H_N=-\frac{H \phi_N^2}{2 M_{\mathrm{pl}}^2}
    \label{eq:friedmann_f_N}
\end{eqnarray}
With a given potential, we can solve the coupled background equation Eqs.(\ref{eq:KG_f_N}) and (\ref{eq:friedmann_f_N}) with the proper initial conditions. The end of inflation can be computed using the slow-roll parameter $\epsilon =1$, and the minimum value of e-folds $N$ constrains the initial value of $\phi$; the initial value for $\phi_N$ and $H$ are obtained from the same background equation.

\subsection{Scalar and tensor perturbations} 

It is standard practice to compute the equations governing the dynamics of linear perturbations in terms of the curvature perturbation $\mathcal{R}$ and tensor perturbation $\psi$. Defining the mode functions for the scalar perturbation $u = -z\mathcal{R}$ and $v_k = a \psi_k$ for the tensor perturbation, with $z$ as the Mukhanov variable $z \equiv a \dot{\phi}/H$. The equation of motion for the Fourier components, $u_k$ and $v_k$, are
\begin{subequations}
\label{eqs:mukhanov_sasaki}
\begin{eqnarray}
    u_k''+\left(k^2-\frac{z''}{z}\right) u_k &=&  0, \label{eq:ms_scalar} \\
    v_k''+\left(k^2-\frac{a''}{a}\right) v_k &=&  0. \label{eq:ms_tensor}
\end{eqnarray}
\end{subequations}
The derivatives in the last equations are with respect to conformal time, and each mode depends on the value of the perturbations wavenumber, $k$. However, given the exponential evolution in inflation and to facilitate the numerical solution, it is convenient to solve Eqs.\eqref{eqs:mukhanov_sasaki} in terms of the number of e-folds, $N$, the equation for the scalar perturbations are rewritten as
\begin{eqnarray}
    u_{\N\N}+\left(1+\frac{H_{\N}}{H}\right) u_{\N}+\left(k^2-\frac{z^{\prime \prime}}{z}\right) \frac{u}{a^2 H^2}=0
\end{eqnarray} \label{eq:MS_scalar_f_n}
with
\begin{equation}
    \frac{z^{\prime \prime}}{z}=a^2 H^2\left(2-\frac{5 H_{\N}}{H}-2\left(\frac{H_{\N}}{H}\right)^2-4 \frac{H_{\N} \phi_{\N\N}}{H \phi_{\N}}-\frac{1}{H^2} \frac{\partial^2 V(\phi)}{\partial \phi^2}\right) \; ,
\end{equation}
for the sake of notation, we have suppressed the dependency of $k$ on $u$ and $v$; the subscript $\N$ denotes differentiation with respect to the number of e-folds variable. The equations for the tensor modes are
\begin{equation}
    v_{\N\N}+\left(1+\frac{H_{\N}}{H}\right) v_{\N}+\left(k^2-\frac{a^{\prime \prime}}{a}\right) \frac{v}{a^2 H^2}=0
\end{equation} \label{eq:MS_tensor_f_n}
with
\begin{equation}
    \frac{a^{\prime \prime}}{a} = H^2 a^2\left(2+\frac{H_{\N}}{H}\right) \; .
\end{equation}

The initial conditions on the perturbations are set when the physical wavelengths are well inside the Hubble radius, i.e., when $k/aH \gg 1$ and the Bunch-Davies vacuum conditions are applied \cite{Bunch:1978yq}. The conditions on both scalar and tensor perturbations are established when $k^2 \gg z''/z$ and $k^2 \gg a''/a$, respectively. In this limit, the $k^2$ term in Eqs.\eqref{eqs:mukhanov_sasaki} dominates, and the modes have an oscillatory solution, $\mathbf{u} \propto \mathrm{exp}\left(-\sfrac{ikN}{aH} \right)$ for a $\mathbf{u} = \{u_k, v_k\}$. The initial conditions at $N_i$ are given after computing the normalization factor of the oscillatory solutions
\begin{eqnarray}\label{eq:initial_conditions}
    \mathbf{u}(N_i) = \frac{1}{\sqrt{2k}} \quad \mathrm{ and } \quad 
    \mathbf{u}_\N(N_i) = -i \sqrt{\frac{k}{2}} \frac{1}{aH}
\end{eqnarray}

The scalar and the tensor power spectra, $\ps(k)$ and $\pt(k)$, are expressed in terms of the mode functions $(u_k, v_k)$ as follows:
\begin{subequations}\label{eqs:PPS}
\begin{eqnarray}
    \ps(k) &=& \frac{k^3}{2\pi^2} | \mathcal{R} |^2    = \frac{k^3}{2\pi^2} \left\vert \frac{ u_k}{z} \right\vert^2_{\N_{e}} ,\label{eq:PPS_scalar}\\
    \pt(k) &=& \frac{k^3}{2\pi^2} | \mathcal{\psi} |^2 = \frac{k^3}{2\pi^2} \left\vert \frac{ v_k}{a} \right\vert^2_{\N_{e}} .\label{eq:PPS_tensor}
\end{eqnarray}
\end{subequations}

To compare the observable parameters with observational data, we need to compute the scalar spectral tilt $n_{s}$ and the tensor-to-scalar ratio $r$ at the pivotal scale ($k_\star = 0.05 \, \mathrm{Mpc^{-1}}$) which are related to the primordial power spectrum as  
\begin{equation}
n_{s}-1\equiv \frac{d\ln \ps}{d\ln k} \,,\quad r=\frac{\pt}{\ps} \,,
\end{equation}

Our aim is to iteratively compute the evolution of the Mukhanov-Sasaki equations for different $k$-modes, Eqs.\eqref{eqs:mukhanov_sasaki}, for the inflation potential with a step described in the section below, first, compute the primordial power spectrum, and then to compute the spectral distortions described in the next section. Numerically, we take the initial conditions when $k/aH \approx 10^2$ for each perturbation mode, which is sufficient enough for the numerical evolution \cite{Adams:2001vc, Mortonson:2009qv, Hazra:2012yn, PhysRevResearch.2.013030, Ragavendra:2020sop}. With the solutions to the background at hand, we can evaluate the coefficients in the equations~\eqref{eqs:mukhanov_sasaki} governing the perturbations. Starting with the initial conditions~\eqref{eq:initial_conditions}, we use method {\it (py)oscode} \cite{PhysRevResearch.2.013030} specialized in fast solutions of oscillatory ODEs to evolve the scalar and tensor perturbations until the on super-Hubble scales, mainly, when $k/aH \approx 10^{-2}$.

\subsection{The step potential} 

Precision measurements using CMB and large-scale structures data have revealed that the density perturbations exhibit a nearly scale-invariant PPS \cite{Wang:2022nml, Sohn:2022jsm, Planck:2018jri, Palma2018}. This observation agrees with canonical inflationary models, which estimate a scale-invariant or Harrison-Zeldovich spectrum \cite{Harrison:1969fb, Zeldovich:1972zz}. Such PPS can be parameterized using a simple power-law. Nevertheless, deviations from the power-law (or features) in the PPS reconstructions might provide a more accurate representation of the observational data \cite{Lodha:2023jru, Hazra:2010ve}; it could also indicate non-trivial and richer inflationary dynamics \cite{Starobinsky:1992ts, Adams:2001vc, Hunt:2004vt, Covi:2006ci, Hamann2010, Chen:2008wn, Hunt2015, Bartolo:2013exa, GallegoCadavid:2014jac, GallegoCadavid:2016wcz}. Distinguishing between statistical fluctuations, noise, systematics in the data, and genuine features derived from the inflaton mechanism remains challenging \cite{Palma2018, Hazra2014, Hunt2015, Hu2014, Aslanyan2014, Hunt2014, Verde2008, Leach2006, TocchiniValentini2005, Mukherjee:2003cz}, and certainly, it will be an interesting avenue to explore with future surveys measuring the SD and getting constraints to the PPS \cite{Chluba:2016bvg, Cabass:2018jgj, Chluba:2019kpb, Clesse:2014pna}.

\begin{figure}[t]
    \centering
    \includegraphics{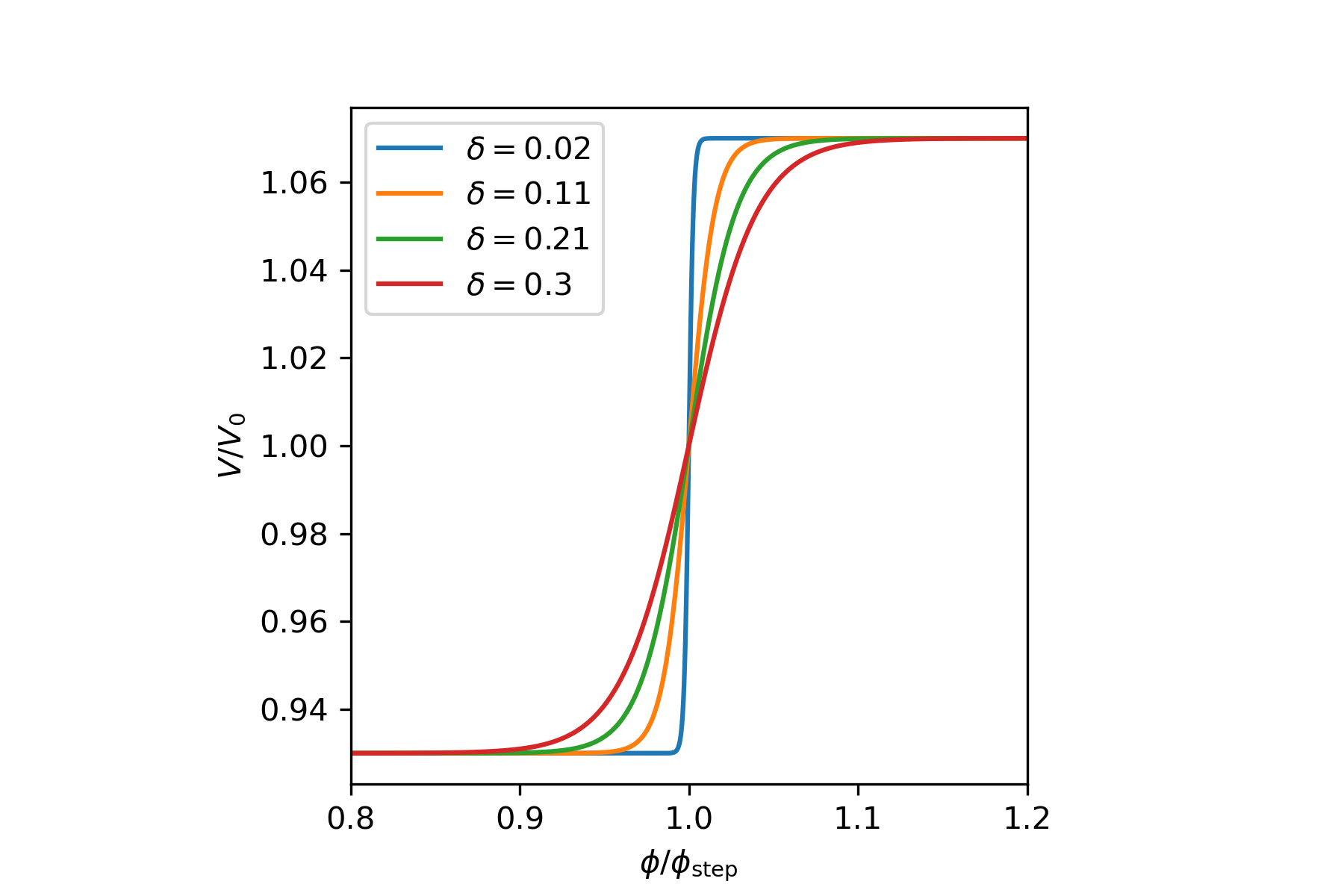}
    \caption{ Inflationary potential. $V_0$ represents the power-law inflation with a power of $n=2/3$. The blue, orange, green and red solid line correspond to step potential (see Eq. \ref{eq:potential}) for fixed $\delta= \{ 0.02,\, 0.11,\, 0.21,\, 0.3\}$ in ${\rm M_{pl}}$ units, respectively, we take $\phi_\text{step}=7.425 \, {\rm M_{pl}}$ and $\beta=0.075$.}
    \label{fig:beta_potential}
\end{figure}

The features in the PPS are intriguing because they cannot be produced by the standard slow-roll inflationary models, and such a departure could be attributed to a step in the inflationary potential in the following way
\begin{equation}
    V(\phi) = M^4\left( \frac{\phi}{\mathrm{M_{pl}}} \right)^n \left[1+\beta\,\tanh\left(\frac{\phi-\phi_\textnormal{step}}{\delta}\right)\right],
    \label{eq:potential}
\end{equation}
where the step occurs at $\phi=\phi_\textnormal{step}$, $M$ is the inflaton mass in Planck mass units ($M_{pl}$); the parameters $\beta$ and $\delta$ denote the amplitude and width of the step, respectively; and $n$ is the power dependency with respect to the field, see Fig.~\ref{fig:beta_potential}. 

Considering the power-law potential alone, $V(\phi) \propto \phi^n$, the case for $n = 2$ is known as the chaotic potential. Cases when $n > 1$, can lead to large--field inflationary models; this is, they can reach super--Planckian distance in field space during inflation and predict large values for the tensor-to-scalar ratio. On the other hand, powers with $n < 1$ lead to small--field models that predict a smaller tensor-to-scalar ratio, which is in agreement with the latest observational upper bounds with $r < 0.037$ using Planck+Bicep-Keck+BAO+lensing data \cite{BICEP:2021xfz, Campeti:2022vom}; where BAO stands for baryon acoustic oscillations. We are motivated to study a fractional power-law with $n=2/3$ because it can lead to predicting a scalar spectral index and a tensor-to-scalar ratio in agreement with current cosmological observations. Additionally, it naturally arrives in effective field theories or in the context of string theory in D-branes scenarios or fields coupled to non-standard gravity theories \cite{Silverstein:2008sg, Nakayama:2010sk, Kallosh:2013yoa, Ashoorioon:2009wa, McAllister:2008hb, McAllister:2014mpa}.

Therefore, considering adding a step to the power-law potential with $n=2/3$ is of significant interest because of its implications for various phases in the evolution of the Universe, from inflation to reheating \cite{Dai:2014jja, Creminelli:2014fca, Planck:2018vyg}. It can also provide insights into the dynamics of scalar fields and their interactions with other fields or particles, and it can be related to dip/enhancements in the primordial power spectrum, which can eventually be constrained through SD observations and does not necessitate super-Planckian field values during inflation, thereby sidestepping potential issues with high-energy physics and quantum gravity effects.

\section{Spectral Distortions} \label{sec:spectral_distortions}  

Spectral distortions are classified depending on their spectral shape and are directly related to the thermodynamic history of the photons. At $z \gtrsim 2 \times 10^6$, Compton scattering is the dominant collision process; it efficiently drives any perturbation to thermodynamic equilibrium, keeping the CMB close to a black body spectrum. Between redshift ranges $5\times 10^4 \lesssim z \lesssim 2\times 10^6$, processes that alter the number-density of photons, such as Bremsstrahlung and double Compton scattering, become inefficient due to the expansion of the Universe. This results in a Bose-Einstein distribution with a non-zero chemical potential $\mu$ for the photons which is approximately constant. Although $\mu$ is largely constant, it is intrinsically a function of both frequency and time \cite{Tashiro:2014pga}. For redshifts $z \lesssim 5 \times 10^4$, Compton scattering becomes inefficient. At this point, background electrons at a higher temperature can boost CMB photons out of equilibrium to create $y$ SD. This effect is conceptually similar to the Sunyaev–Zeldovich effect but is pertinent to the early Universe \cite{SunyaevZel1969}.

Observable SD is expected on scales smaller than those for galaxies, restricting the amplitude of initial perturbations on small scales and complementing the results from CMB anisotropy observations, where constraints on smaller scales are influenced by the Silk damping effect. Consequently, this provides constraints for the initial power spectrum at higher wavenumbers, $1 \lesssim k \lesssim 10^4 \; \mathrm{Mpc}^{-1}$. Precise measurements of the CMB power spectrum along with SD constraints from COBE/FIRAS observations~\cite{Mather:1993ij, Fixsen:1996nj} established upper bounds for $\mu \lesssim 9\times 10^{-5}$ and $y \lesssim 1.5 \times 10^{-5}$ at $2\sigma$ C.L. Moreover, the ARCADE 2 experiment provided slightly more conservative bounds with $\mu < 6\times 10^{-4}$ at $2\sigma$ C.L. ~\cite{Arcade2_experiment}. The TRIS experiment found tighter constraints, $\mu \lesssim 6\times 10^{-5}$, for frequencies close to $\nu \simeq 1$ GHz~\cite{Gervasi:2008eb}. The absence of primordial black holes and ultracompact minihalos has helped set upper bounds on the small-scale primordial power spectrum amplitude. Specifically, on the amplitude of the primordial power spectrum to be less than $0.01-0.06$ over for the wavenumber range $0.01 \lesssim k \lesssim 10^{23} \; {\rm Mpc^{-1}}$. Forthcoming projects, like PIXIE and its extensions, promise to explore scales that were previously out of reach~\cite{Chluba:2019nxa, Kogut:2019vqh}. These advances will shed further light on the inflationary epoch and necessitate a reevaluation of our theoretical predictions.

In this work, we employ an analytic approach for small spectral distortions $\mu$ and $y$ ($\mu$, $y \ll 1$). During the radiation-dominated epoch, energy stored in small-scale density perturbations gets dissipated through photon diffusion, the Silk damping process. This dissipation led to the heating of the CMB photons, which in turn caused $\mu$ and $y$ SD. The magnitude of the SD directly depends on the shape and amplitude of the PPS of curvature perturbations, $\mathcal{P}_{_{\mathrm{R}}}$, larger perturbations would lead to more significant heating and more pronounced distortions.
An analytic approach that estimates the impact on $\mu$ and $y$ SD from the damping is given by 
\begin{equation}
    \label{eq:sd5}
    i ~\approx~ \int_{k_\mathrm{min}}^{\infty} \frac{k^{2} \dd k}{2 \pi^2} \ps(k) \, W^{i} (k)\,,
\end{equation}
where $k_\mathrm{min} = 1~\text{Mpc}^{-1}$, and $W^i(k)$ represents the efficiency at which the acoustic damping and thermalization effects contribute at a given scale ($k$-modes) to the $i$--SD, where $i\in\{y,\mu\}$. For instance, the function peaks at certain redshifts where the distortions are most efficiently produced and diminishes at others. The window function is defined by
\begin{subequations}
\begin{eqnarray}
    W^{y}(k) 
    &\approx& \frac{C^{2}}{2} \,\mathrm{e}^{-k^{2} / 32^{2}}\,,\\
    W^{\mu} (k) ~&\approx&~ 2.8\, C^{2}  \exp{ \left(  -\frac{\left[ \frac{k}{1360 ~k_\mathrm{min}} \right]^{2} }{1 + \left[ \frac{k}{260~k_\mathrm{min}} \right]^{0.3} + \left[ \frac{k}{340~k_\mathrm{min}} \right]} \right)} - 5.6\,  W^{y}(k)\,.
\end{eqnarray}
\end{subequations}
where the factor $C$ is the amplitude of the perturbation; it depends on the type of the perturbations. Specifically, for adiabatic modes, is a constant $C\approx0.902$. The exponential term introduces the Silk damping, which account for the k-modes at which perturbations get damped due to photon diffusion. Therefore, the window function ensures that small-scale perturbations (high-$k$ values) are suppressed in their contribution to the distortions~\cite{Chluba:2012gq,Chluba:2013dna}. The code \texttt{CosmoTherm} solves the cosmological perturbation equations to obtain accurate results for the energy release rates caused by the damping of the acoustic modes \cite{Chluba:2013dna}. Comparing its outcomes with those from the analytic approach reveals an excellent consistency. The differences are mainly noticeable for the y--SD, but is less than 10\% difference \cite{Chluba:2016bvg}. 

%
The total difference of the photon intensity spectrum compared to the Planck distribution, $\Delta I$, can be computed based on the photon phase-space distribution, the photon Boltzmann equation, and Green's function approach~\cite{Lucca:2019rxf, Fu:2020wkq}. The Green's functions translate an energy injection at a certain $k$-mode (or redshift) to a distortion of a specific frequency observed at the current time. The total intensity spectrum in the presence of these distortions can be computed as,
\begin{equation}\label{eq:sd1}
\Delta I (x)
= \frac{\Delta T}{T} G(x) + y\, Y(x)  + \mu\, M(x)\,.
\end{equation}
The $G(x)$, $M(x)$, and $Y(x)$ are normalized spectral-shape functions that represent the different types of spectral distortions. The first term is a shift in the temperature of the CMB spectrum, $\Delta T$, and $G(x)$ is defined as
\begin{equation}
    G(x)= -x \frac{\partial B(x)}{\partial x}  \,,
    \qquad\text{with}\quad
    B(x)=\frac{1}{\left(e^{x}-1\right)}
    \quad\text{and}\quad
    x \equiv \frac{h \nu}{k_{B} T}\,,  
\end{equation}
where $x$ is the dimensionless frequency variable, $h$ is the Planck constant, $\nu$ is the frequency, $k_B$ is the Boltzmann constant, and $T$ is the photon temperature.
$M(x)$ and $Y(x)$ function accompanying the $\mu$ and $y$ spectral distortions, respectively, are defined as
\begin{subequations}
\begin{eqnarray}
    Y(x) &=& G(x)\left(x\frac{e^x+1}{e^x-1}-4 \right) \,, \\
    M(x) &=& G(x)\left(\alpha_\mu - \frac{1}{x} \right)\,, 
\end{eqnarray}
\end{subequations}
where the coefficient $\alpha_\mu$ is a correction factor, determined under the constraint that the $\mu$--SD conserves the photon number density. Its approximate value is $\alpha_\mu = 0.4561$. The term $ \nicefrac{\Delta T}{T}$ in Eq.~\eqref{eq:sd1} does not introduce any distortion to the blackbody spectrum. Instead, It corresponds to a departure of the average temperature of the CMB from the blackbody temperature. Such a deviation is challenging to detect ~\cite{Lucca:2019rxf}. In this context, we focus on computing the $y$ and $\mu$ SD. Earlier research has demonstrated that SD can set stringent limits on the amplitude of the primordial power spectrum, especially when derived from certain inflationary potentials that exhibit distinct features \cite{Chluba:2013wsa,Chluba:2013dna,Chluba:2012we,Chluba:2012gq}. Our study specifically delves into the SD generated by the inflationary models characterized by a step potential.

\section{Results}\label{sec:results}  
We study the power-law inflation potential with a power of $n=2/3$ that can match current cosmological observations and stand out in various theoretical models. We incorporated a step in the potential (see Eq.\eqref{eq:potential}) because of its various cosmological implications, from inflationary phases to reheating. This extension offers insights into dips or enhancements in the PPS that, on the one hand, can be related to inflationary dynamics and, on the other hand, can be constrained by upcoming experimental missions, e.g., PIXIE, via the measurement of SD of the CMB photons. Using Eq.\eqref{eq:sd5} we compute the amplitude of the $\mu$ and $y$ SD type, which allows us to forecast the intensity signal through Eq.\eqref{eq:sd1} given by the step-potential model. 

The potential is defined by three main free parameters: the scale when the step occurs $\phi_{\text{step}}$, the amplitude of the step $\beta$, and the duration of the step $\delta$. We also fix the inflaton mass $M\approx 2.66\times 10^{-4} \, {\rm M_{pl}}$ to avoid degeneracy with the amplitude. For comparison with observations, we established the pivot scale at $k_\star=0.05\text{ Mpc}^{-1}$, where the amplitude of scalar perturbations is $\ln(10^{10}A_\mathcal{S})= 3.040\pm 0.016$ and the tensor-to-scalar ratio is $r < 0.37$. \cite{Planck:2018jri, BICEP:2021xfz}.

\begin{figure}[t!]
     \centering
     \begin{subfigure}{0.328\linewidth}
     \centering
         \includegraphics[width=\linewidth]{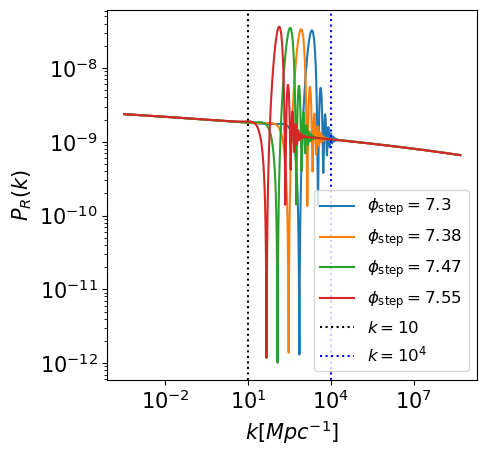}
     \caption{$\beta$ and $\delta$ fixed.}
     \end{subfigure}
     \hfill
     \begin{subfigure}{0.328\linewidth}
     \centering
         \includegraphics[width=\linewidth]{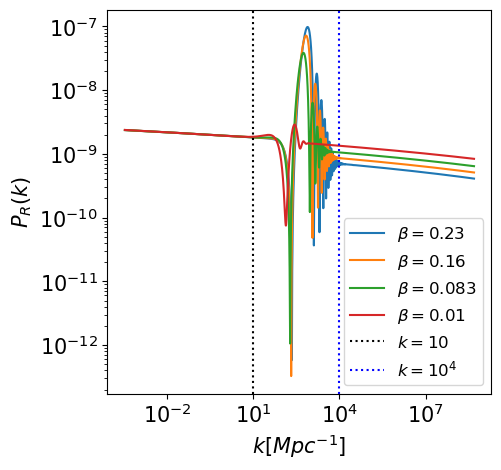}
         \caption{$\phi_\text{step}$ and $\delta$ fixed.}
     \end{subfigure}
     \hfill
     \begin{subfigure}{0.328\linewidth}
     \centering
         \includegraphics[width=\linewidth]{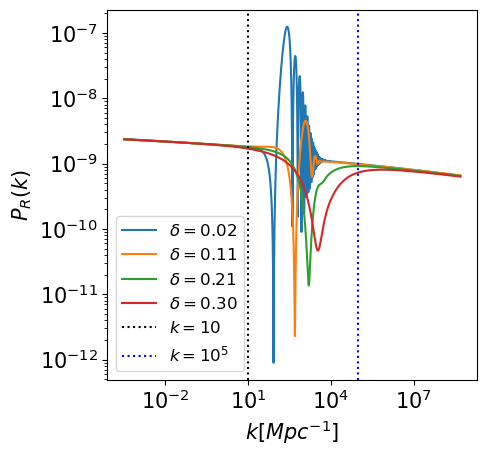}
         \caption{$\phi_\text{step}$ and $\beta$ fixed.}
     \end{subfigure}
     \hfill
     
     \caption{Scalar primordial power spectrum, $\ps(k)$, of inflationary models based in a step potential, computed via Eq.\eqref{eq:PPS_scalar} for $\phi_\text{step}$ (left), $\beta$ (middle) and $\delta${-dependence} (right). The dotted vertical lines separate the regions where $\ps(k)$ is almost scale invariant of the regions where is scale variant. In (a) $\beta=0.075$ and $\delta=0.055\, {\rm M_{pl}}$, in (b) $\phi_\text{step}=7.425\, {\rm M_{pl}}$ and $\delta=0.055\, {\rm M_{pl}}$, and in (c) $\phi_\text{step}=7.425\, {\rm M_{pl}}$ and $\beta=0.075$. }
     \label{fig:mu_SPPS} 
\end{figure}
First, we solve the background evolution defined by the Klein-Gordon (Eq.\eqref{eq:KG_f_N}) and Friedmann (Eq.\eqref{eq:friedmann_f_N}) equations. Then, we solve the Mukhanov-Sasaki equation to obtain the evolution of each of the scalar and tensor modes with Eqs.\eqref{eq:MS_scalar_f_n} and \eqref{eq:MS_tensor_f_n}, respectively. When the evolution of each mode is stretched beyond the causal horizon (when $k \gg (aH)^{-1}$), their amplitude stops evolving; after the end of inflation, the modes re-enter the horizon with the same amplitude, start evolving, and eventually lead to the formation of galaxies, clusters, and other cosmic structures. The frozen squared amplitude of the scalar (and tensor) modes at the horizon crossing defines the scalar (tensor) PPS; see Eqs.\eqref{eqs:PPS}.

In Fig.~\ref{fig:mu_SPPS}, we show examples of the PPS that we obtained; the introduction of a step in the power-law potential leads to oscillations in both scalar and tensor PPS. The characteristics of these oscillations are determined by the three free parameters $\phi_{\text{step}}$, $\beta$, and $\delta$ of the potential. $\phi_\text{step}$ shifts the scales where the oscillations are located; for larger values of $\phi_\text{step}$, the oscillations begin at smaller wavenumber values, $k$. The $\beta$ parameters change the initial amplitude and the number of oscillations; the amplitude is small, and a few oscillations are found for a low value of $\beta$. The $\delta$ parameter affects the duration and number of the oscillations; it is also degenerated with $\beta$ and also affects the amplitude of the oscillations. The parameters $\beta$ and $\delta$ have a direct impact on the dynamics of the inflaton field and the expansion ratio of the Universe; for instance, large values of $\beta$ combined with small $\delta$ values could temporarily stop the Universe from inflating since $\epsilon > 1$ can be achieved for certain parameter combination. High resolution is required for small values of $\delta$ in order to compute the high oscillatory features.

\begin{figure}[t]
     \centering
     \begin{subfigure}[b]{0.325\textwidth}
     \centering
         \includegraphics[width=\textwidth]{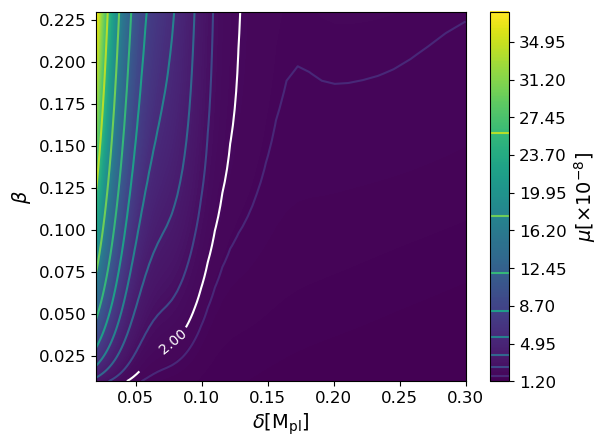}
         \caption{$\phi_\text{step}=7.30 {\, \rm M_{pl}}$}
         \label{fig:mu_heatmap_a}
     \end{subfigure}
     \hfill
     \begin{subfigure}[b]{0.325\textwidth}
     \centering
         \includegraphics[width=\textwidth]{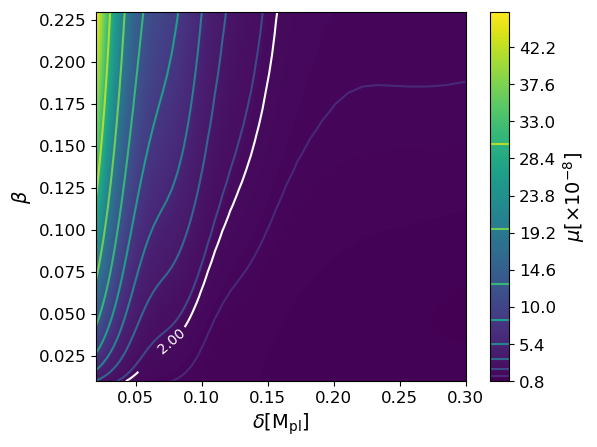}
         \caption{$\phi_\text{step}=7.43 {\, \rm M_{pl}}$}
         \label{fig:mu_heatmap_b}
     \end{subfigure}
     \hfill
     \begin{subfigure}[b]{0.325\textwidth}
     \centering
         \includegraphics[width=\textwidth]{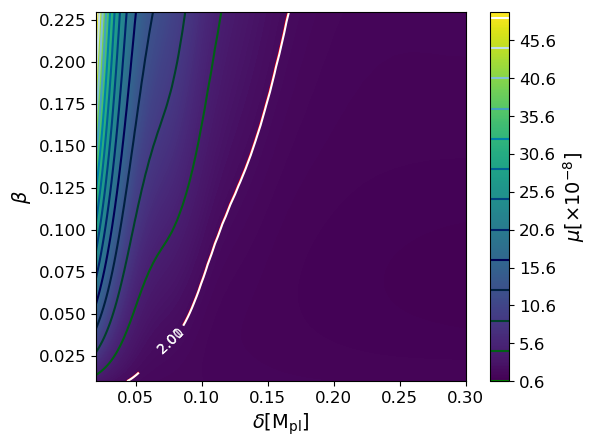}
         \caption{$\phi_\text{step}=7.55 {\, \rm M_{pl}}$}
         \label{fig:mu_heatmap_c}
     \end{subfigure}
     \hfill
     \begin{subfigure}[b]{0.325\textwidth}
     \centering
         \includegraphics[width=\textwidth]{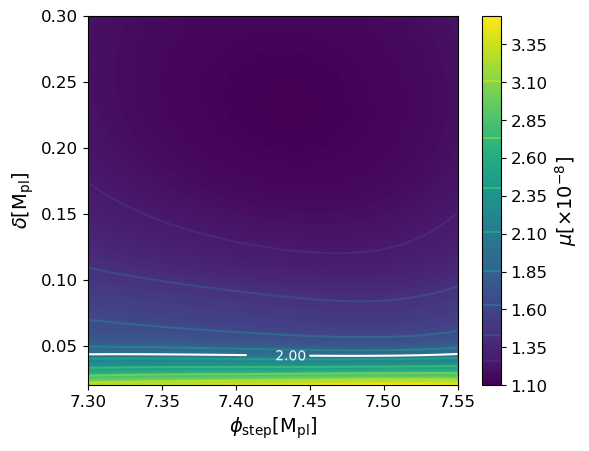}
         \caption{$\beta=0.01$}
         \label{fig:mu_heatmap_d}
     \end{subfigure}
     \hfill
     \begin{subfigure}[b]{0.325\textwidth}
     \centering
         \includegraphics[width=\textwidth]{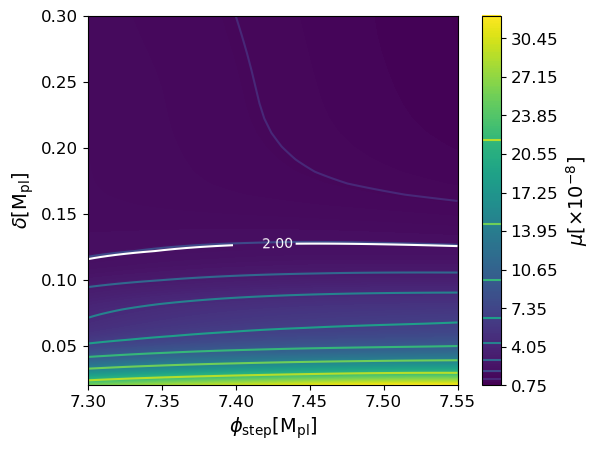}
         \caption{$\beta=0.12$}
         \label{fig:mu_heatmap_e}
     \end{subfigure}
     \hfill
     \begin{subfigure}[b]{0.325\textwidth}
     \centering
         \includegraphics[width=\textwidth]{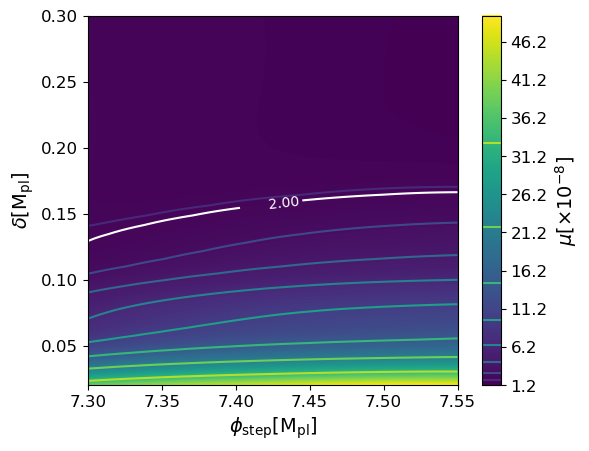}
         \caption{$\beta=0.23$}
         \label{fig:mu_heatmap_f}
     \end{subfigure}
      \hfill

     \begin{subfigure}[b]{0.325\textwidth}
     \centering
         \includegraphics[width=\textwidth]{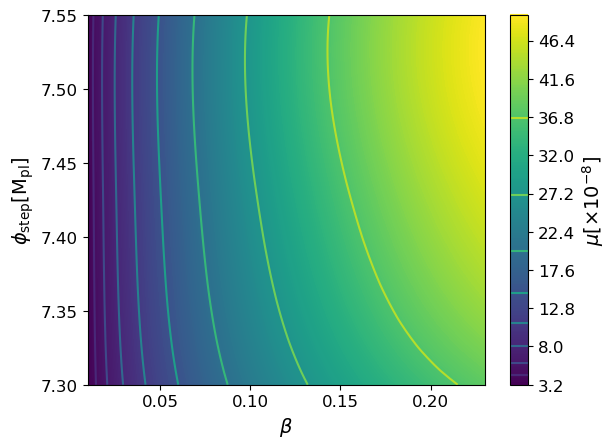}
         \caption{$\delta=0.02 {\, \rm M_{pl}}$}
         \label{fig:mu_heatmap_g}
     \end{subfigure}
     \hfill
     \begin{subfigure}[b]{0.325\textwidth}
     \centering
         \includegraphics[width=\textwidth]{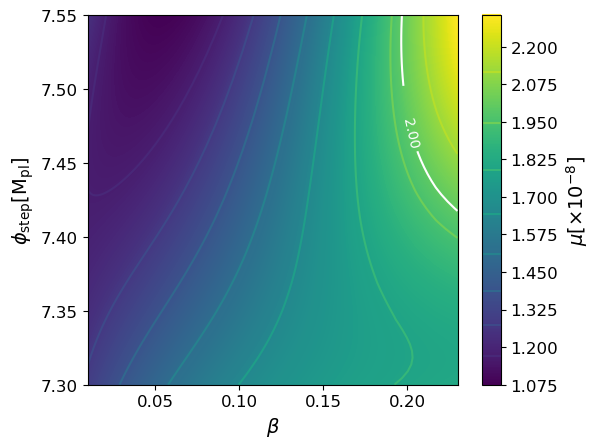}
         \caption{$\delta=0.16 {\, \rm M_{pl}}$}
         \label{fig:mu_heatmap_h}
     \end{subfigure}
     \hfill
     \begin{subfigure}[b]{0.325\textwidth}
     \centering
         \includegraphics[width=\textwidth]{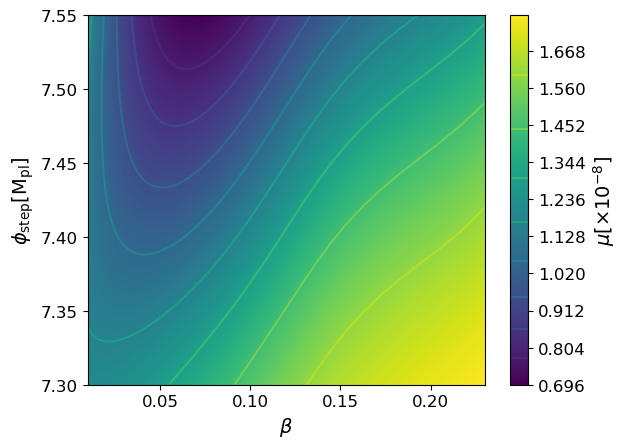}
         \caption{$\delta=0.30 {\, \rm M_{pl}}$}
         \label{fig:mu_heatmap_i}
     \end{subfigure}
     \caption{The color-map shows the computations of $\mu$--SD with Eq.\eqref{eq:sd5}, based on the step--potential inflationary model Eq.\eqref{eq:potential}. When present, white line denotes the value predicted by the $2/3$-power-law ($\beta = 0$) for which $\mu_{\rm 2/3}=2.0113\times10^{-8}$.    }
     \label{mu_heatmap}
\end{figure}

Despite the fact that some studies claim oscillation in the CMB observational window $10^{-3} \leq  k \leq 10^{-1} \, \text{Mpc}^{-1}$  our focus is on estimating the SD at smaller scales. Specifically, we study at wavenumbers ranging from $1 \leq  k \leq 10^{4} \, \text{Mpc}^{-1}$; these scales will likely be covered in upcoming SD observational surveys. To guarantee that the oscillatory features fall within the SD observational window of these future surveys and to ensure that inflation is uninterrupted by the step (by the condition $\epsilon < 1$ for the duration of inflation), the parameter space needs to be initially constrained by
\begin{align*}
    7.3 \leq \frac{\phi_\text{step}}{{\rm M_{pl}}} &\leq 7.55, & 0.01 \leq \beta &\leq 0.23, & \text{and} &\hspace{1cm} 0.02 \leq \frac{\delta}{{\rm M_{pl}}} \leq 0.3.
\end{align*}

After the perturbation modes re-enter the cosmological horizon, they start to evolve; however, the acoustic modes with a wavelength shorter than the mean free path of photons face a damping behavior (known as Silk-damping), observed at high-$l$ values in the CMB angular power spectrum. The energy released due to the Silk damping is dissipated into the monopole of the radiation field, which creates SD. The CMB monopole is directly connected to the primordial power spectrum via perturbation theory. Therefore, by measuring $\mu$ and $y$ SD, using Eq.\eqref{eq:sd5}, we indirectly can provide information related to inflation through the PPS. 

In Fig.~\ref{mu_heatmap}, we present the forecast for the $\mu$--SD. The white curve, when visible, represents the value predicted by the $2/3$-power-law model; this is, with $\beta=0$. For this case, the model predicts a value for $\mu_{\nicefrac{2}{3}} = 2.0113\times10^{-8}$ which also coincides for $n=2/3$ for the axion monodromy model in \cite{Henriquez-Ortiz:2022ulz}; this prediction is also proximate to the PIXIE detection threshold at a $2\sigma$ C.L. and the prediction of the $\Lambda$CDM model, which is $\mu_{\rm \Lambda CDM}=2.00\times10^{-8}$ \cite{Kogut:2011xw, Chluba:2012we, Chluba:2016bvg}, making it the canonical value. The top of Fig.~\ref{mu_heatmap} show the $\beta-\delta$ parameter space for specific fixed values of $\phi_\text{step} = \{7.3, ~7.43, ~7.55\}$ in  Planck mass units, ${\rm M_{pl}}$. The middle section focuses on the $\delta-\phi_\text{step}$ parameter space, holding $\beta = \{0.01, ~0.12, ~0.23\}$. The bottom section shows the $\phi_\text{step}-\beta$ parameter space, with fixed value of $\delta = \{0.02, ~0.16, ~0.3\}$ in ${\rm M_{pl}}$ units.

Variations in the $\beta$ and $\delta$ parameters influence $\mu$--SD more than those in $\phi_{\rm step}$. The amplitude arising from the step in the PPS is primarily influenced by the $\beta$ parameter; an increase in amplitude corresponds to a larger $\mu$ value. Oscillations in the PPS become more pronounced as $\delta$ decreases and less pronounced as $\delta$ increases. There is a threshold in the $\delta-\beta$ parameter space in which the $\mu$ values fall below the detection threshold, and it is for $\delta > 0.17 \, {\rm M_{pl}}$ that will be outside the observational sensitivity for any combination of $\mu$ and $\beta$, see Figs.~\ref{fig:mu_heatmap_a}, \ref{fig:mu_heatmap_b} and \ref{fig:mu_heatmap_c}. The sensitivity of $\mu$ to $\phi_\text{step}$ is lower compared to the other parameters because $\phi_\text{step}$ mainly shifts the oscillation in the PPS; for instance, the maximum value of $\mu$ changes 23\% comparing the two extremes of the $\phi_{\rm step}$ parameter, see Figs.~\ref{fig:mu_heatmap_a} and \ref{fig:mu_heatmap_c}. Moreover, when computing the $\mu$--SD over a broad wavenumber range, its value remains largely unaffected by $\phi_\text{step}$, see Figs.~\ref{fig:mu_heatmap_d}, \ref{fig:mu_heatmap_e}, and \ref{fig:mu_heatmap_f}.

When $\phi_\text{step}$ is held constant, the magnitude of the $\mu$--SD is most pronounced at the highest $\beta$ values, driven primarily by the oscillation amplitudes. Consequently, a decrease in $\beta$ leads to smaller values for $\mu$. The peak value of $\mu$--SD ($\mu = 49.6 \times 10^{-8}$) is an order of magnitude larger than the fiducial model, results from the parameter set: $\phi_\text{step}=7.53\, {\rm M_{pl}}$, $\beta=0.23$, and $\delta=0.02 \, {\rm M_{pl}}$ (see Figs.~\ref{fig:mu_heatmap_c}, \ref{fig:mu_heatmap_f}, and \ref{fig:mu_heatmap_g}). This combination, with a clearly high $\beta$ and a low $\delta$, ensures pronounced oscillations in the PPS. In contrast, the minimal $\mu$ value ($\mu = 0.70 \times 10^{-8}$) is 34\% lower than the fiducial model, and results from $\phi_\text{step}=7.55\, {\rm M_{pl}}$, $\beta=0.07$, and $\delta=0.3\, {\rm M_{pl}}$ (see Figs.~\ref{fig:mu_heatmap_c} and \ref{fig:mu_heatmap_i}), a set characterized by a small $\beta$ and large $\delta$ values.

\begin{figure}[t]
     \centering
     \begin{subfigure}[b]{0.325\textwidth}
     \centering
         \includegraphics[width=\textwidth]{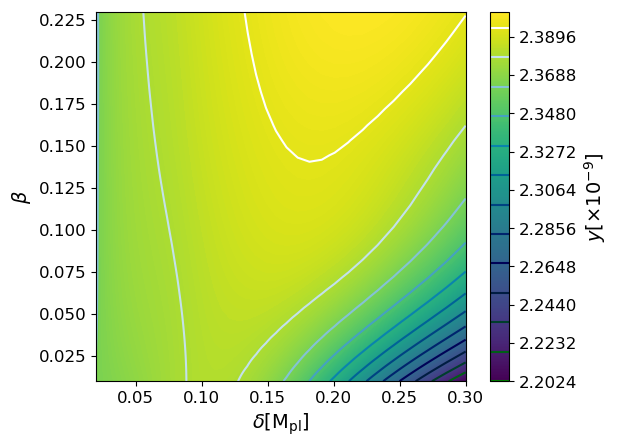}
         \caption{$\phi_\text{step}=7.30 {\, \rm M_{pl}}$}
         \label{fig:y_heatmap_a}
     \end{subfigure}
     \hfill
     \begin{subfigure}[b]{0.325\textwidth}
     \centering
         \includegraphics[width=\textwidth]{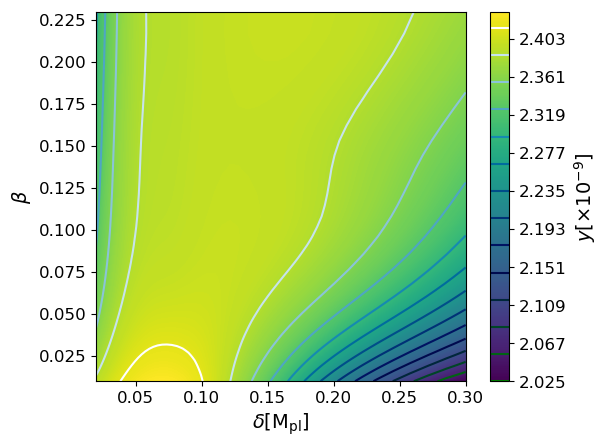}
         \caption{$\phi_\text{step}=7.43 {\, \rm M_{pl}}$}
         \label{fig:y_heatmap_b}
     \end{subfigure}
     \hfill
     \begin{subfigure}[b]{0.325\textwidth}
     \centering
         \includegraphics[width=\textwidth]{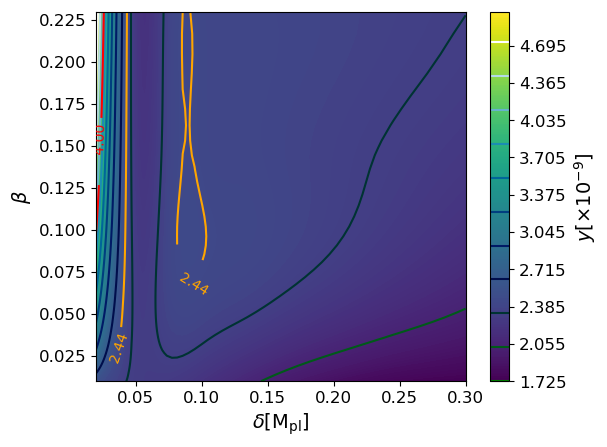}
         \caption{$\phi_\text{step}=7.55 {\, \rm M_{pl}}$}
         \label{fig:y_heatmap_c}
     \end{subfigure}
     \hfill
     \begin{subfigure}[b]{0.32\textwidth}
     \centering
         \includegraphics[width=\textwidth]{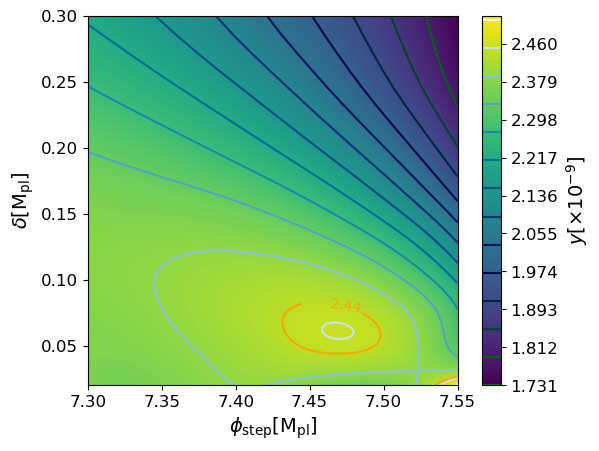}
         \caption{$\beta=0.01$}
         \label{fig:y_heatmap_d}
     \end{subfigure}
     \hfill
     \begin{subfigure}[b]{0.32\textwidth}
     \centering
         \includegraphics[width=\textwidth]{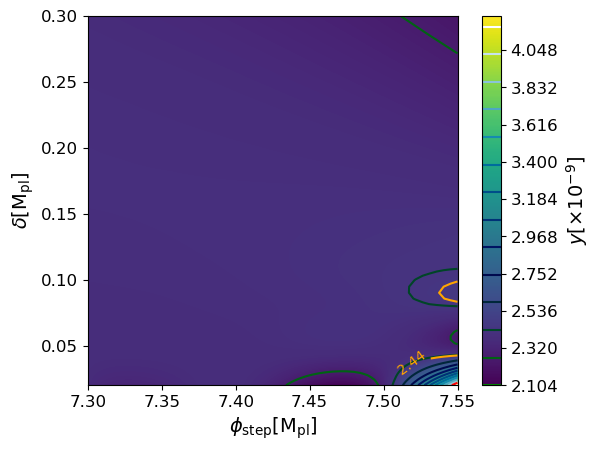}
         \caption{$\beta=0.12$}
         \label{fig:y_heatmap_e}
     \end{subfigure}
     \hfill
     \begin{subfigure}[b]{0.32\textwidth}
     \centering
         \includegraphics[width=\textwidth]{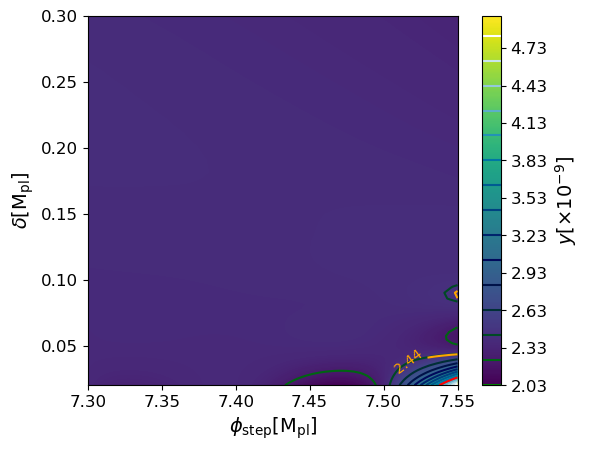}
         \caption{$\beta=0.23$}
         \label{fig:y_heatmap_f}
     \end{subfigure}
      \hfill      
     \begin{subfigure}[b]{0.32\textwidth}
     \centering
         \includegraphics[width=\textwidth]{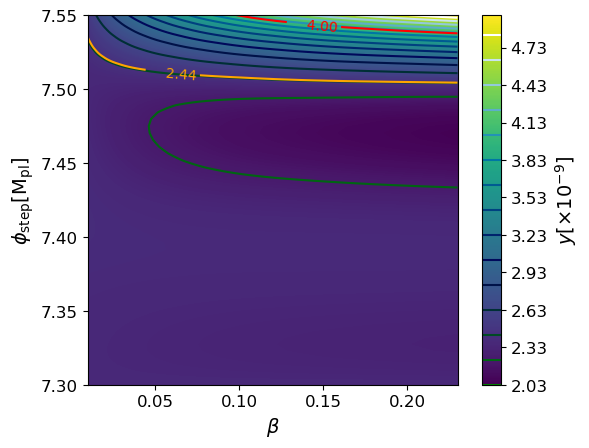}
         \caption{$\delta=0.02 {\, \rm M_{pl}}$}
         \label{fig:y_heatmap_g}
     \end{subfigure}
     \hfill
     \begin{subfigure}[b]{0.32\textwidth}
     \centering
         \includegraphics[width=\textwidth]{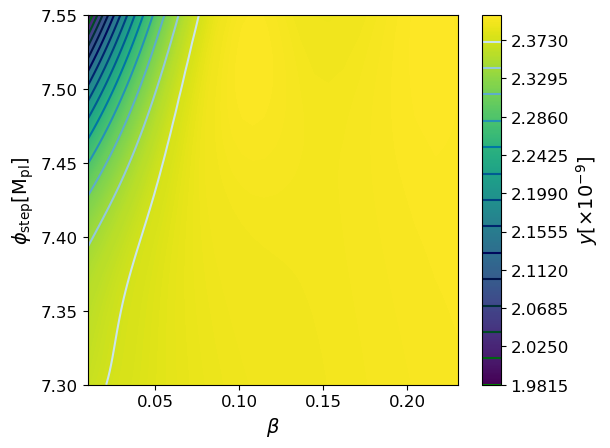}
         \caption{$\delta=0.16 {\, \rm M_{pl}}$}
         \label{fig:y_heatmap_h}
     \end{subfigure}
     \hfill
     \begin{subfigure}[b]{0.32\textwidth}
     \centering
         \includegraphics[width=\textwidth]{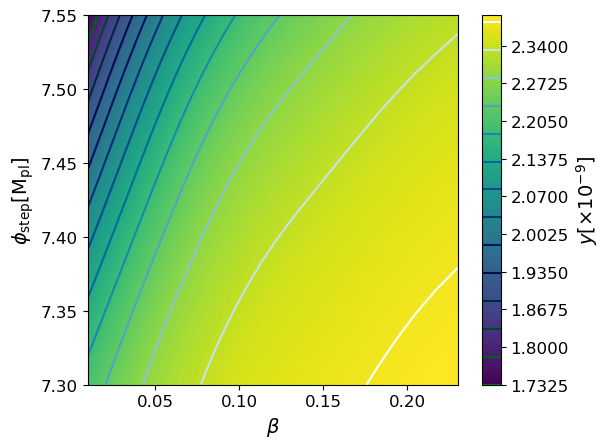}
         \caption{$\delta=0.30 {\, \rm M_{pl}}$}
         \label{fig:y_heatmap_i}
     \end{subfigure}
     \caption{The color-map shows the computations of $y$--SD with Eq. \eqref{eq:sd5}, based on the step potential inflationary model Eq.\eqref{eq:potential}. When present, the orange line denotes the value predicted by the $2/3$-power-law ($\beta = 0$) for which $y =2.4376\times 10^{-9}$ \cite{Henriquez-Ortiz:2022ulz}. The red line represents the observational threshold of PIXIE mission at the $2\sigma$ C.L., corresponding to $y=4\times 10^{-9}$ \cite{Chluba:2012we}. }
     \label{y_heatmap}
\end{figure}

In Fig.~\ref{y_heatmap}, color maps are presented to illustrate the forecasted value of the $y$--type SD, analogous to what was done for $\mu$. The orange curve, when visible, indicates the value as predicted by the $2/3$-power-law. Specifically, for $\beta = 0$, the $y$-SD is $y_{\nicefrac{2}{3}} = 2.44\times 10^{-9}$, which also coincides with the power-law behavior of the axion monodromy model \cite{Henriquez-Ortiz:2022ulz}, which is 7\% lower in value in contrast to the $y_{\rm \Lambda CDM} = 3.54 \times 10^{-9}$ for the $\Lambda$CDM vanilla model \cite{Chluba:2016bvg}. When depicted, a red line marks the observational threshold for PIXIE at a $2\sigma$ C.L., which is $y = 4\times 10^{-9}$ \cite{Chluba:2012we}.

In the parameter space, a specific region yields $y$-type SD values that exceed the observational sensitivity. This region is characterized by the conditions: $\phi_\text{step} > 7.54 \, {\rm M_{pl}}$, $\beta > 0.10$, and $\delta < 0.026\, {\rm M_{pl}}$. Additionally, another region marked by $\phi_\text{step} > 7.51 \, {\rm M_{pl}}$ and $\delta < 0.03\, {\rm M_{pl}}$, and independent of the variation with respect to the $\beta$ parameter, produces $y$--SD values that surpass the expected $2/3$-power-law value. These regions are shown in Figs.~\ref{fig:y_heatmap_c}, \ref{fig:y_heatmap_e}, \ref{fig:y_heatmap_f}, and \ref{fig:y_heatmap_g}.

The sensitivity of the $y$--SD to variations in $\phi_{\rm step}$ is greater than that of the $\mu$--SD, however, we can see a shorter spectrum for $y$ values when $\delta > 0.1 \, {\rm M_{pl}}$, as seen in Figs.~\ref{fig:y_heatmap_h} and \ref{fig:y_heatmap_i}. The largest values for $y$--SD arise from a combination of large $\phi_{\rm step}$ and $\beta$ values coupled with small $\delta$ values. Specifically, the peak value forecasted for $y$--SD is $y = 4.99 \times 10^{-9}$, which is twice the value associated with the fiducial $2/3$-power-law, $y = 2.05 \, y_{\nicefrac{2}{3}}$, and 41\% larger than $y_{\rm \Lambda CDM}$. This peak corresponds to the parameter set $\phi_\text{step}=7.55\, {\rm M_{pl}}$, $\beta=0.23$, and $\delta=0.02\, {\rm M_{pl}}$, as seen in Figs.~\ref{fig:y_heatmap_c}, \ref{fig:y_heatmap_f}, and \ref{fig:y_heatmap_g}. Conversely, the minimal $y$--SD value is $y = 1.73 \times 10^{-9}$, which is 30\% lower than the expected value for a step-less potential, $y_{\nicefrac{2}{3}}$. This minimum is obtained with the parameters $\phi_\text{step}=7.55\, {\rm M_{pl}}$, $\beta=0.1$, and $\delta=0.3\, {\rm M_{pl}}$, seen in Figs.~\ref{fig:y_heatmap_c}, \ref{fig:y_heatmap_d}, and \ref{fig:y_heatmap_i}. It is interesting to notice that maximum and minimum values of both $\mu$ and $y$ SD are when $\phi_{\rm step} \approx 7.55$ ${\rm M_{pl}}$, the largest value for $\phi_{\rm step}$ in our parameter space; conversely, $\phi_{\rm step} \approx 7.30 \, {\rm M_{pl}}$ is closer to the $\nicefrac{2}{3}$-power-law and is the minimum value in the explored parameter space.

\begin{table}[t]
    \centering    
    \begin{tabular}{l|c|c|c|c}
     & \multicolumn{2}{c}{$\mu/10^{-8}$} & \multicolumn{2}{|c}{$y/10^{-9}$} \\
    \hline
    Parameters & max = 49.58 & min = 0.70 & max = 4.99 & min = 1.73 \\
    \hline \hline
    $\phi_{\rm step} \; [{\rm M_{pl}} ]$   & 7.53  & 7.55  & 7.55    & 7.55 \\
    $\beta$                 & 0.23  & 0.07  & 0.23    & 0.10 \\
    $\delta \; [{\rm M_{pl}}]$   & 0.02  & 0.30   & 0.02    & 0.30 \\
    \end{tabular}
    \caption{ Maximum (depicted as max) and minimum (depicted as min) values of the spectral distortions $\mu$ and $y$ computed for the step potential in the parameter space \{$\phi_{\rm step},~\beta,~\delta$\}.}
    \label{tab:max_and_min}
\end{table}

Furthermore, knowing the range of $\mu$ and $y$ predicted by our model in the parameter space used is important to analyze whether the features of the step are distinguishable from the $\nicefrac{2}{3}$-power-law and $\Lambda CDM$. Both the maximum and minimum values obtained from the analysis for the $\mu$ and $y$ SD by the step-potential model, along with the parameter space to obtain such values, are presented in Table \ref{tab:max_and_min}. 

\begin{figure}[t]
     \centering
     
     \begin{subfigure}[b]{0.325\textwidth}
     \centering
         \includegraphics[width=\textwidth]{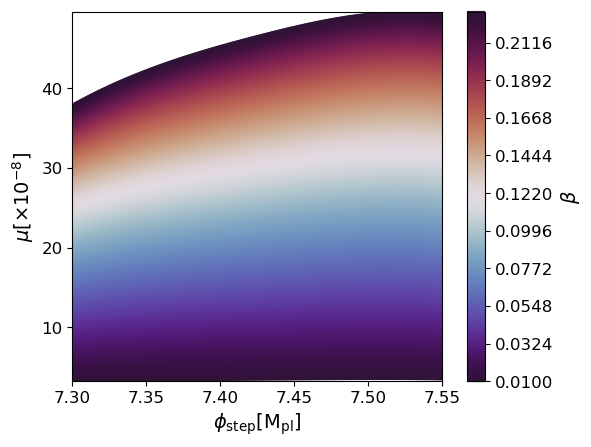}
         \caption{$\delta=0.02 \, {\rm M_{pl}}$}                
         \label{mu_in_vp_a}
     \end{subfigure}
     \hfill
     \begin{subfigure}[b]{0.325\textwidth}
     \centering
         \includegraphics[width=\textwidth]{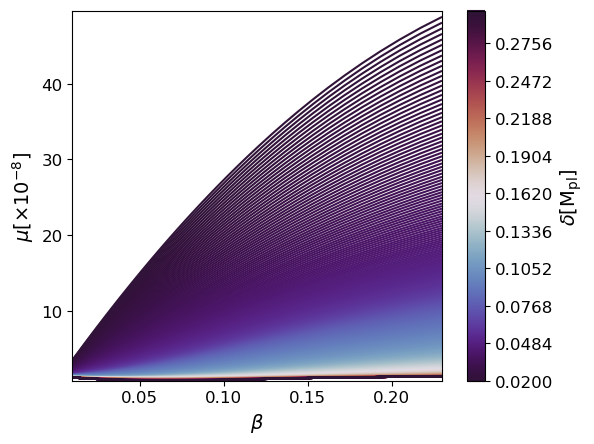}
         \caption{$\phi_\text{step}=7.53 \, {\rm M_{pl}}$}         
         \label{mu_in_vp_b}
     \end{subfigure}
     \hfill
     \begin{subfigure}[b]{0.32\textwidth}
     \centering
         \includegraphics[width=\textwidth]{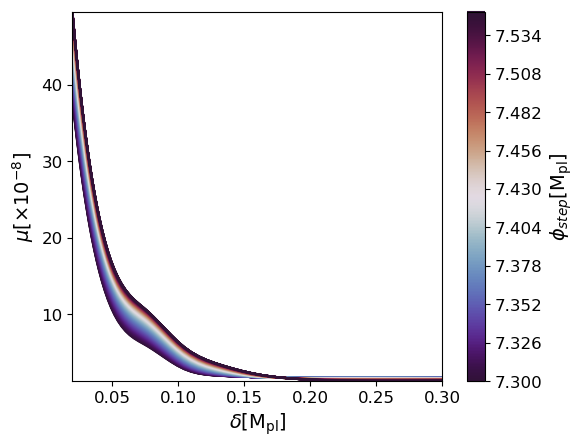}
         \caption{$\beta=0.23$}
         \label{mu_in_vp_c}
     \end{subfigure}
     \hfill
     \begin{subfigure}[b]{0.325\textwidth}
     \centering
         \includegraphics[width=\textwidth]{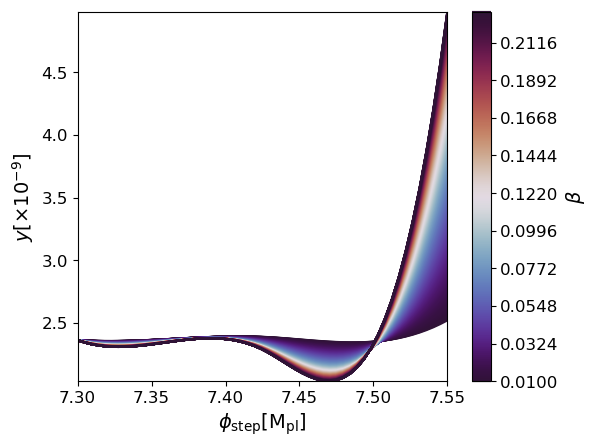}
         \caption{$\delta=0.02 \, {\rm M_{pl}}$}         
         \label{mu_in_vp_d}
     \end{subfigure}
     \hfill
     \begin{subfigure}[b]{0.32\textwidth}
     \centering
         \includegraphics[width=\textwidth]{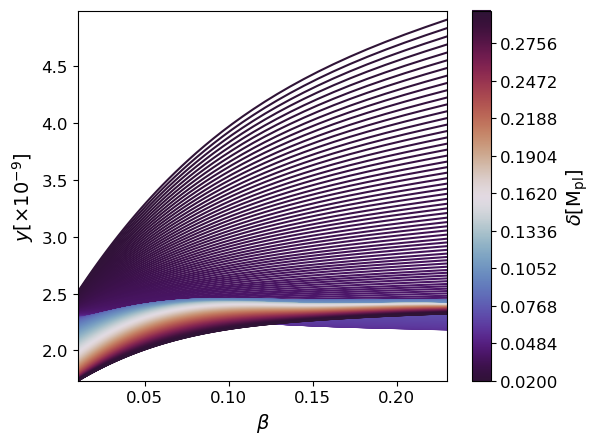}
         \caption{$\phi_\text{step}=7.55 \, {\rm M_{pl}}$}         
         \label{mu_in_vp_e}
     \end{subfigure}
    \hfill
    \begin{subfigure}[b]{0.32\textwidth}
     \centering
         \includegraphics[width=\textwidth]{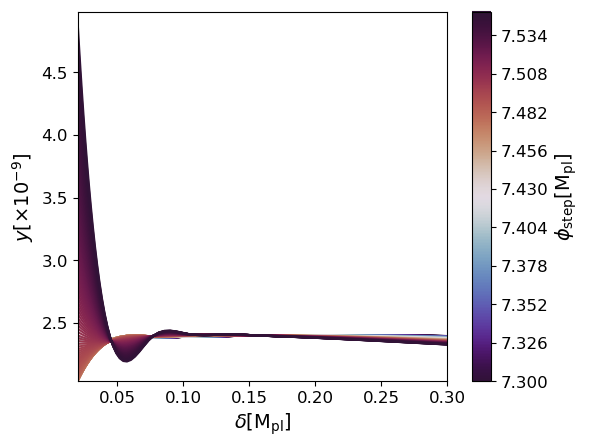}
         \caption{$\beta=0.23$}
         \label{mu_in_vp_f}
    \end{subfigure}
    \caption{We present marginalized computations for the spectral distortions $\mu$ (shown at the top) and $y$ (displayed at the bottom). On the left side, distortions are plotted as a function of $\phi_\text{step}$ for different values of $\beta$, with a fixed value of $\delta=0.02 \, {\rm M_{pl}}$. In the middle are the distortions as a function of $\beta$ for different values of $\delta$. For those plots, $\phi_\text{step} = 7.53\, {\rm M_{pl}}$ remains constant for $\mu$ and $\phi_\text{step} = 7.55\, {\rm M_{pl}}$ for $y$. On the right side, distortions are plotted as a function of $\delta$ for a range of $\phi_\text{step}$ values, with a fixed value of $\beta = 0.23$. The fixed value of the parameters is set to yield the maximum distortion value. }
     \label{mu_in_vp}
\end{figure}

To gain a better understanding of the connection between the SD and the inflation potential, we plotted the marginalization over one specific parameter in Fig.~\ref{mu_in_vp}. The top plots show the marginalization used to compute $\mu$--SD, while the bottom plots are for $y$--SD. Examining Fig.~\ref{mu_in_vp} from left to right: in \ref{mu_in_vp_a} and \ref{mu_in_vp_d}, we showcase the dependency of $\phi_\text{step}$, with variation in $\beta$ and a fixed $\delta = 0.02\, {\rm M_{pl}}$; in \ref{mu_in_vp_b} and \ref{mu_in_vp_e}, we depict the $\beta$ dependency while varying $\delta$ and holding $\phi_\text{step} = 7.53 \, {\rm M_{pl}}$ ($\phi_\text{step} = 7.55 \, {\rm M_{pl}}$) for $\mu$--SD ($y$--SD) . Lastly, \ref{mu_in_vp_c} and \ref{mu_in_vp_f} presents the $\delta$ dependency, with variations in $\phi_\text{step}$ and a fixed $\beta = 0.23$. We selected these specific fixed values because they yield the maximal values for both $\mu$ or $y$ SD, see Table \ref{tab:max_and_min}.

Figures \ref{mu_in_vp_b} and \ref{mu_in_vp_e} show that as the value of $\beta$ increases --indicating a larger step in the PPS-- both $\mu$ and $y$ SD become large. Regarding the $\delta$ variable, large distortions are obtained at smaller $\delta$ values because the oscillation in the PPS emerges more abruptly. Thus, for large $\delta$ values, the step transition becomes smoother, and distortions quickly converge to the value given by the $2/3$-power-law potential, see Figs.~\ref{mu_in_vp_c} and \ref{mu_in_vp_f}. A marginalized analysis reveals that the $y$--SD values exhibit a pattern reminding of damping oscillation with both increasing-$\delta$ values and decreasing-$\phi_{\rm step}$ values. This behavior is corroborated by the degenerate region in the $\delta$-$\beta$ parameter space as depicted in Fig.~\ref{fig:y_heatmap_c}, which corresponds to the oscillations observed in Fig.~\ref{mu_in_vp_f} for the range $0.07 < \delta / {\rm M_{pl}} < 0.11$. The minimum of the $y$--SD value in Fig.~\ref{mu_in_vp_d} is related to the circular or partially-circular isolines in Figs.\ref{fig:y_heatmap_b}, \ref{fig:y_heatmap_d}, and \ref{fig:y_heatmap_g}. From these figures, one can discern the parameter space producing a $y$-SD value below the value expected from the $2/3$-power-law, which is defined by the combination of parameters $\beta < 0.024$, $7.44 \lesssim \phi_{\rm step} /  {\rm M_{pl}} \lesssim 7.49$, and $0.05 \lesssim \delta /  {\rm M_{pl}}\lesssim 0.08$ 

Several interesting observations can be made from the marginalized plots for $\mu$ and $y$ SD. Referring to Figs.~\ref{mu_in_vp_c} and \ref{mu_in_vp_d}, when $\phi_{\rm step} < 7.40 \, {\rm M_{pl}}$, the values for $\mu$ fall within the range $39.0 \lesssim \mu/10^{-8} \lesssim 45.3$. In contrast, the $y$--SD approaches the value defined by the $2/3$-power-law, $y \approx y_{\nicefrac{2}{3}}$. A similar behavior is evident in Figs.~\ref{mu_in_vp_c} and \ref{mu_in_vp_f} for the $\delta$ interval $0.10 \lesssim \delta / {\rm M_{pl}} \lesssim 0.19$. In this range, the $\mu$-SD is $1 < \mu/\mu_{\nicefrac{2}{3}} < 2.85 $, while the $y$ distortion remains close to $y \approx y_{\nicefrac{2}{3}}$. These unique patterns could lead to distinct features in the photon intensity that can distinguish between the step-less potential or the $\Lambda$CDM model; also see below the discussion for the G and H cases when computing the intensity.

We identify a distinct region in which small values of $\beta < 0.075$ and large values of $\delta \gtrsim 0.12 \, {\rm M_{pl}} $ lead to obtain values $y/y_{\nicefrac{2}{3}} < 1$ when we expected a ratio close to one, as seen in Fig.~\ref{mu_in_vp_e}. The reason for this behavior can be attributed to the features of the PPS in which for $\beta < 0.075$ induces a half-slow oscillation, which looks like a dip in the PPS rather than having multiple oscillations, see Fig.~\ref{fig:poso_pps} in the appendix. This, along with the behavior of the correspondent window function, we obtain $y$-SD smaller than the expected for $2/3$-power-law, in contrast to the $\mu$-SD, which tends to the fiducial value. This can also be seen in Fig.~\ref{fig:mu_heatmap_a}, in which the region of interest we obtain similar values of $\mu$; comparing now with Fig.~\ref{fig:y_heatmap_a} in which a clear relation between $\beta$ and $\delta$ is shown within the isolines tending to values of $y$ which are up to $9.7\%$ smaller than $y_{\sfrac{2}{3}}$.

\begin{figure}[t]
     \centering
\includegraphics[width=0.45\textwidth]{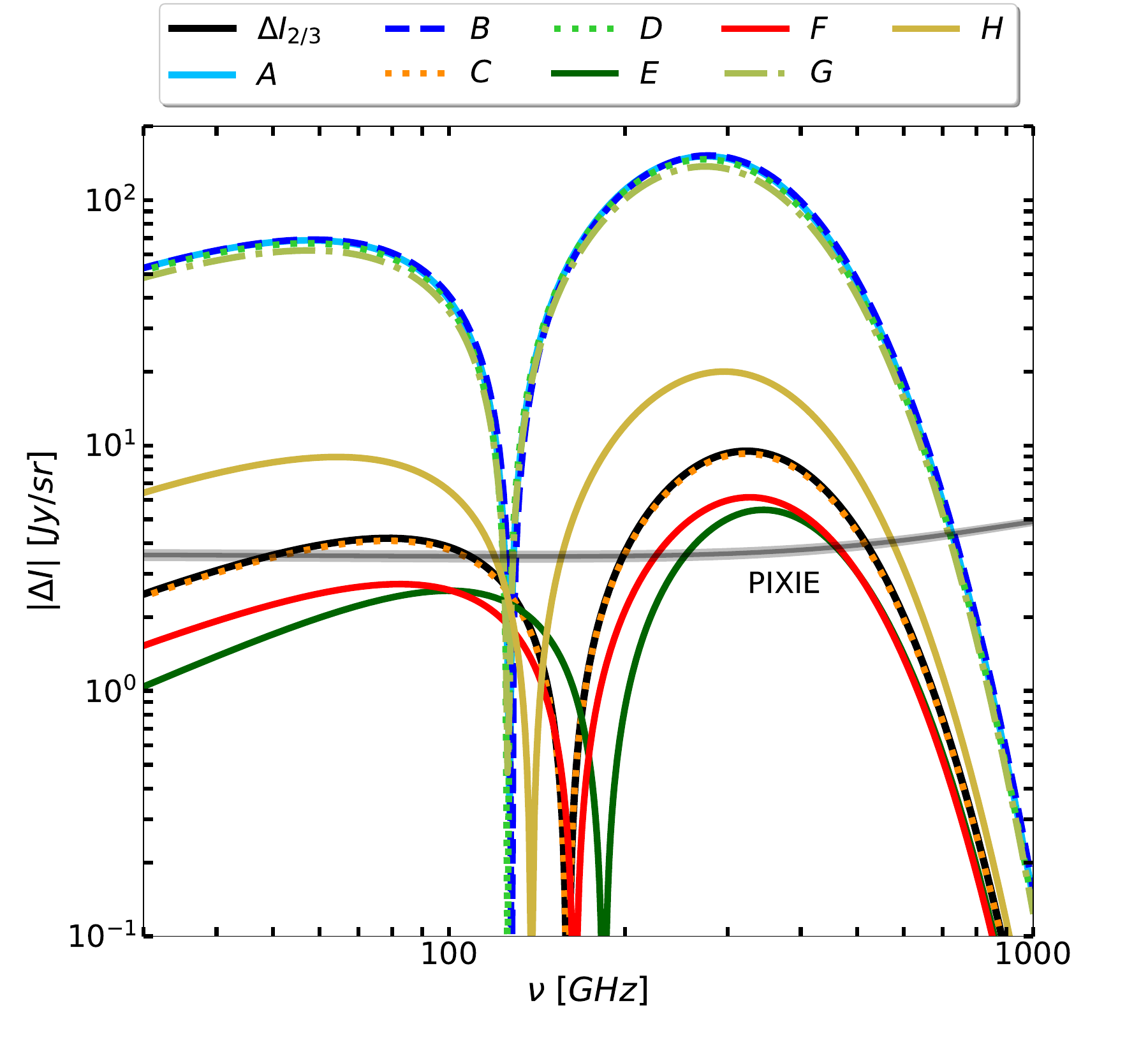}
\includegraphics[width=0.45\textwidth]{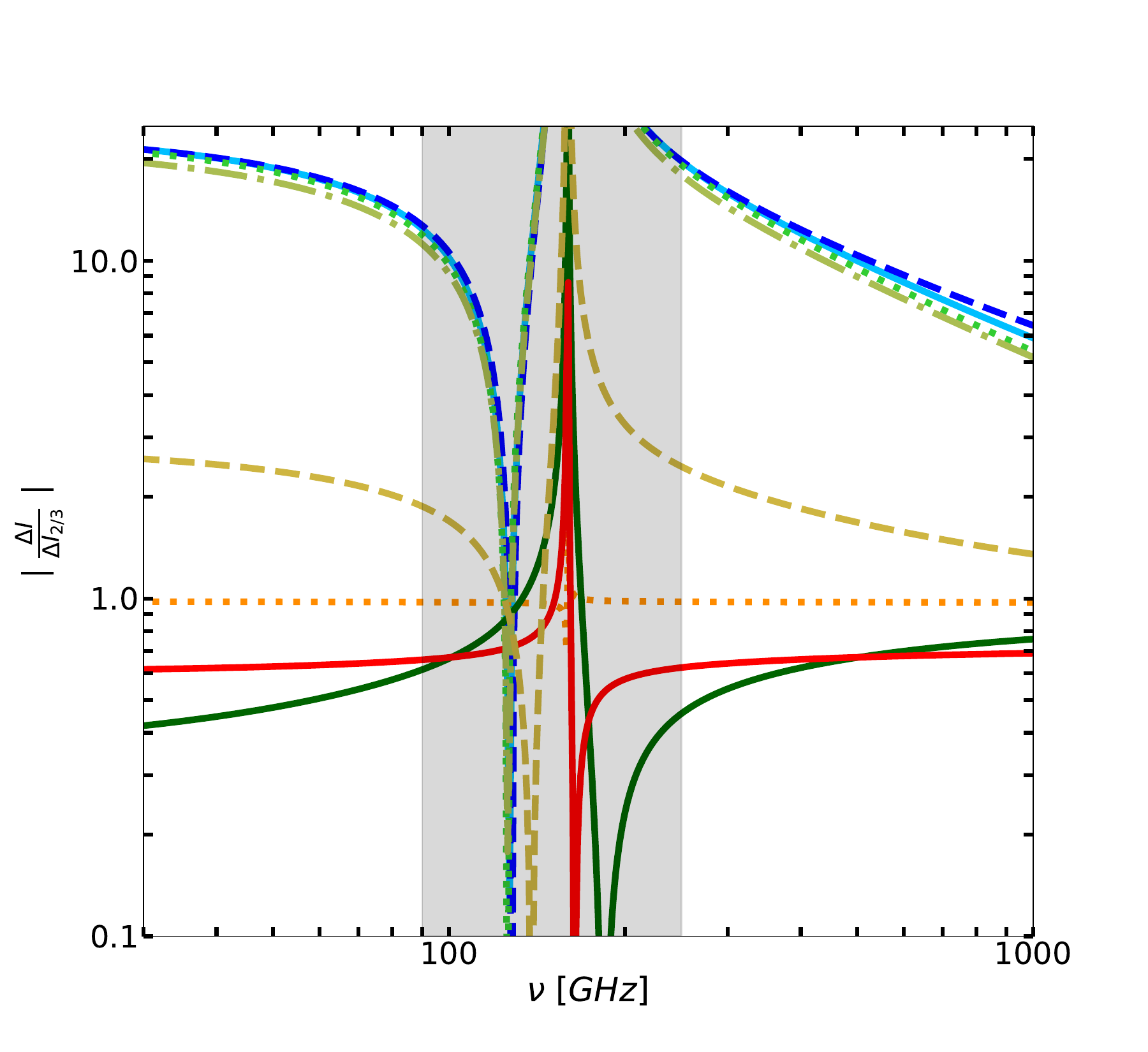}
\caption{Predictions on the contribution to the distortion $\Delta I$ of the photon intensity arising from $\mu$ and $y$ SD in the chaotic with a step inflationary models is show in the left panel for the different cases shown in Table.~\ref{tab:intensity}: four solid curves correspond to A (cyan), E (dark green), F (red), and H (olive); the $\Delta I_{2/3}$ prediction (determined from $\mu_{2/3}$ and $y_{2/3}$) is the solid black line. Three dotted curves are shown: C (orange), D (green), and G (yellow-green). The grey line is the predicted PIXIE sensitivity at a $2\sigma$ C.L. The plot on the right panel shows the ratio of the intensities computed with Eq.\eqref{eq:sd1} and the $\nicefrac{2}{3}$-power-law prediction, $\left\langle  \Delta I/\Delta I_{\nicefrac{2}{3}} \right\rangle $, for the same scenarios as in the left panel and keeping the same labels. In Table~\ref{tab:intensity}, the average difference is computed for frequencies outside the range $100\,\mathrm{GHz}\lesssim\nu\lesssim250\,\mathrm{GHz}$ (the grey region).}
\label{plot:sd_di}
\end{figure}

We compute the contributions of $\mu$ and $y$ to the distortions of the photon intensity spectrum $\Delta I (\nu)$, see Eq.\eqref{eq:sd1}, and compare the results with $\Delta I_{\nicefrac{2}{3}}$, that was obtained from the contribution of $\mu_{2/3}$ and $y_{2/3}$ (similar is with the signal $\Delta I_{\rm \Lambda CDM} $, related by  $\mu_{\rm \Lambda CDM}$ and $y_{\rm \Lambda CDM}$), and to the sensitivity of the PIXIE experiment. In the left panel of Fig.~\ref{plot:sd_di} we have plotted the intensity in units of $\mathrm{Jy}/\mathrm{sr}=10^{-26}\, \mathrm{W}\,\mathrm{m}^{-2}\mathrm{Hz}^{-1}\mathrm{sr}^{-1}$ for the potential with a step.

Now, let us choose some configurations in the parameter space to compute the photon intensity spectrum. Table~\ref{tab:intensity} and Fig.~\ref{plot:sd_di} show the cases chosen; below, we explain the reasons for those points. First, we choose the points that give the maximum of $\mu$--SD and $y$--SD, characterized by having large $\beta$ and small $\delta$ values, both depicted as points A and B, respectively. C label is the point closer to $2/3$-power-law; this is characterized by small $\beta$ and large $\delta$. Point D describes the minimum value on the local minimum for $y$--SD shown in Fig.~\ref{fig:y_heatmap_d} in the range of $7.41 < \phi_{\rm step}/{\rm M_{pl}} < 7.50$. Points E and F depicted the points with the lowest $\mu$ and $y$ SD, respectively. Point G and H have a $y \approx y_{2/3}$ but $\mu \neq \mu_{2/3}$. 

For cases having high oscillatory features, we notice that the resolution, although important when computing the PPS, does not significantly affect the computation of the intensity, mainly because the oscillation transit between each oscillation very rapidly does not affect the integral computation in Eq.~\eqref{eq:sd5}, having approximately $300$ data points per oscillation was enough for the intensity computation.

\begin{table}[t]
    \centering
    \begin{tabular}{|r c c c r r c c l l|}\hline
     Label & $\phi_{\rm step}/{\rm M_{pl}}$ & $\beta$ & $\delta/{\rm M_{pl}}$ & $\mu/10^{-8}$ & $\mu/\mu_{2/3}$ & $y/10^{-9}$ &
     $y/y_{2/3}$ & $\left\langle \frac{\Delta I}{\Delta I_{\nicefrac{2}{3}}}  \right\rangle $ & $\left\langle  \frac{\Delta I}{\Delta I_{\rm \Lambda CDM}}  \right\rangle $  \\ \hline
     A & $7.53$ & $0.23$ & $0.02$ & $49.58$ & $24.60$ & $3.31$ & $1.36$ & $11.08$ & $3.81$  \\
     B & $7.55$ & $0.23$ & $0.02$ & $49.34$ & $24.54$ & $4.98$ & $2.04$ & $11.40$ & $3.86$  \\
     C & $7.30$ & $0.01$ & $0.05$ & $1.97$  & $0.97$  & $2.37$ & $0.97$ & $0.97$  & $0.25$ \\
     D & $7.48$ & $0.23$ & $0.02$ & $48.63$ & $24.18$ & $2.04$ & $0.84$ & $10.60$ & $3.68$  \\
     E & $7.55$ & $0.07$ & $0.30$ & $0.70$  & $0.35$  & $2.09$ & $0.86$ & $0.77$ & $0.14$ \\
     F & $7.55$ & $0.01$ & $0.30$ & $1.21$  & $0.60$  & $1.73$ & $0.71$ & $0.68$ & $0.16$ \\
     G & $7.40$ & $0.23$ & $0.02$ & $45.31$ & $22.53$ & $2.37$ & $0.97$ & $9.98$ & $3.45$ \\
     H & $7.55$ & $0.23$ & $0.11$ & $5.74$  & $2.85$  & $2.41$ & $0.99$ & $1.74$ & $0.53$ \\ \hline
    \end{tabular}
    \caption{Contribution arising from $\mu$ and $y$ SD  that were used for calculate with Eq.~\eqref{eq:sd1} the distortion $\Delta I$ of the photon intensity. $\Delta I_{\nicefrac{2}{3}} $ was obtained from the contribution of $\mu_{2/3}=2.011 \times 10^{-8}$ and $y_{2/3}=2.438 \times 10^{-9}$, similar with the signal $\Delta I_{\rm \Lambda CDM} $ calculated from  $\mu_{\rm \Lambda CDM}=2.00 \times 10^{-8}$ and $y_{\rm \Lambda CDM}=3.54 \times 10^{-9}$\cite[table 3]{Henriquez-Ortiz:2022ulz}.}
    \label{tab:intensity}
\end{table}

When the ratio $y/y_{\nicefrac{2}{3}}$ is larger than $\mu/\mu_{\nicefrac{2}{3}}$ the intensity signal tends to shift to larger frequency values; see E and F cases in Fig.~\ref{plot:sd_di}. Other cases fulfill that $\mu/\mu_{\nicefrac{2}{3}}$ is larger than $y/y_{\nicefrac{2}{3}}$, in which case the intensity curve shifts to smaller frequency values. In cases A, B, D, and G, we notice that the larger the $\mu$ value, the larger the intensity signal, these four cases have $\mu/\mu_{\nicefrac{2}{3}} \gtrsim 22.53$ and the intensity signal is $\Delta I/ \Delta I_{\nicefrac{2}{3}} \gtrsim 9.98$. For the D case is a local minimum for the $y$--SD; however, $\mu$ dominates when computing the intensity. On the other hand, E and F cases have the lowest $\mu/\mu_{\nicefrac{2}{3}}$ ratio, along with the fact that its $y$--SD values are $y/y_{\nicefrac{2}{3}}<1$, their intensity signal get below the sensitivity of PIXIE, except for a small window at large frequencies. 

The C case is indistinguishable from the $2/3$-power-law, we should mention that this point was found for $\phi_{\rm step} = 7.30\, {\rm M_{pl}}$, though variation in the region $ 7.30 <\phi_{\rm step}/{\rm M_{pl}} \lesssim 7.43$ gives all similar $\mu$ and $y$ values; looking for when the C point deviates from $\mu_{\nicefrac{2}{3}}$ (because $\mu$ variate more rapidly than $y$), we found that, when fixing $\phi_{\rm step}$ and $\beta$ and when $\delta \lesssim 0.04\, {\rm M_{pl}}$ satisfy $\mu/\mu_{\nicefrac{2}{3}} > 1.1$, this is $10\%$ above from the step-less case, see Fig.~\ref{fig:mu_heatmap_a}. Similarly, fixing $\phi_{\rm step}$ and $\delta$ as in the C case but changing $\beta$, we get $\mu/\mu_{\nicefrac{2}{3}} \geq 1.1$ when $\beta > 0.02$; the last two are, therefore, for signals that would be at least 10\% stronger than the $2/3$-power-law, otherwise, the signal would be indistinguishable from the fiducial model or could get below the PIXIE observational threshold, this region also coincides with the parameter space producing a $y$-SD value above the value expected from the $2/3$-power-law discussed above.

The G and H cases were already described; both are cases has a $y/y_{\nicefrac{2}{3}} \approx 1$ and $\mu/\mu_{\nicefrac{2}{3}} > 1$. Though the intensity in the G case is dominated by the large value for the $\mu$--SD, the H case could be more interesting since it represents an intermediate case for the intensity, $\Delta I/\Delta I_{\nicefrac{2}{3}} \approx 1.74$, in which we have a shift of the intensity signal to smaller frequency values that can only be achieved through the step on the potential. 

\section{Conclusion}\label{sec:conclutions}  
We have conducted an analysis of a power-law inflationary potential with a power of $n=2/3$ enriched by a step feature as detailed in Eq.\eqref{eq:potential}. We aimed to compute the PPS and explore the resulting SD and their impact on inflationary dynamics. This work is also motivated considering the increasing capability of experimental missions like PIXIE, which could potentially detect SD in the CMB. Moreover, models with $n = \nicefrac{2}{3}$ are in agreement at 2$\sigma$ C.L. with recent observations, producing a smaller tensor-to-scalar ratio. Furthermore, the step potential emerges in various theoretical frameworks for understanding various cosmological phases, offering insights into scalar field dynamics, their interactions, and potential imprints on cosmological observables.

In analyzing the impact of a step in the inflaton potential, we observed that, unlike smooth potentials, which typically yield monotonically varying power spectra, a step can induce distinct features in the PPS. These include at certain scales oscillations or specific enhancements/suppressions.  If these features fall outside the current observational window, they can be constrained with upcoming observational data, thus offering a robust method to evaluate the feasibility of step-potential inflationary models. The three parameters of the potential drive the features of the oscillations: $\beta$, $\delta$, and $\phi_{\text{step}}$. The parameter $\beta$ influences the amplitude of SD, affecting both $\mu$ and $y$ types; the $\delta$ parameter correlates with the duration and the abruptness of the oscillations, while $\phi_{\rm step}$ determines the onset of these oscillations in the $k$-space of the PPS.  
    
To ensure that the oscillations generated by our model are detectable within the observational window of future surveys, specifically in the range $1 < k < 10^4$ ${\rm Mpc^{-1}}$, we carefully constrained our initial parameter space. This approach also ensured that the inflationary dynamics remained uninterrupted, with $\epsilon < 1$. Our analysis reveals that the constraints on the parameter space vary depending on whether we are examining $\mu$ or $y$ SD. However, it is feasible to analyze each parameter independently. 

Let us start with the parameter $\phi_{\rm step}$. When $\mu/\mu_{\nicefrac{2}{3}}>1$ the value is largely unaffected of variation of $\phi_{\rm step}$ a dependence with $\phi_{\rm step}$ is shown when $\mu/\mu_{\nicefrac{2}{3}} < 1$, although these clearly falls below the observational threshold. In contrast, $y$-SD displays a distinct dependence on $\phi_{\rm step}$ values, with notable variations across different ranges of $\phi_{\rm step}$; going from low to high values: for $7.30 < \phi_{\rm step}/ {\rm M_{pl}} \lesssim 7.43$ the ratio of $y/y_{\nicefrac{2}{3}} \approx 1$, this is, the $y$-SD closely resembles the SD predicted by the $\nicefrac{2}{3}$-power-law model. For  $7.43 < \phi_{\rm step}/{\rm M_{pl}} \lesssim 7.49$ we are in a local minimum for $y$-SD, where most case we have $y/y_{\nicefrac{2}{3}} \lesssim 1$, within this range, the $y$-SD is generally weaker compared to the standard. For $7.49 < \phi_{\rm step}/ {\rm M_{pl}} \lesssim 7.54$, the step in the potential becomes markedly distinct from the typical power-law behavior, and we obtain stronger SD, $y/ y_{\nicefrac{2}{3}} > 1$. Lastly, for $\phi_{\rm step} \gtrsim 7.54 \, {\rm M_{pl}}$ the $y$-SD exceeds the observational threshold of PIXIE.

The $\beta$ parameter scales the magnitude of the $\mu$-SD. Specifically, as the value of $\beta$ increases, so does the $\mu$-SD. Conversely, a decrease in $\beta$ leads to a reduction in $\mu$-SD. As $\beta$ approaches zero, the $\mu$-SD return to the standard $\nicefrac{2}{3}$-power-law model, $\mu/\mu_{\nicefrac{2}{3}} \rightarrow 1$. The $\mu$-SD are heavily influenced by a combination of both $\beta$ and $\delta$, with a lesser dependence on $\phi_{\rm step}$. For instance, setting $\delta = 0.05 \, {\rm M_{pl}}$, when $\beta > 0.02$, $\mu$-SD values rise above the $2/3$-power-law, in particular $\mu/\mu_{\nicefrac{2}{3}} \gtrsim 1.1$. In terms of $y$-SD, the interplay of $\beta$ and $\delta$ also plays a significant role, leading to three distinct scenarios: For $0.01 < \beta \lesssim 0.024$ and $\delta \gtrsim 0.04\, {\rm M_{pl}}$, in this parameter range we observe a counterintuitive phenomenon where $y/y_{\nicefrac{2}{3}} < 1$. This is attributed to the presence of half-oscillations or dips in the PPS, as explained earlier. For $0.024 \lesssim \beta \lesssim 0.1$ and $\phi_{\rm step} \gtrsim 7.51 \, {\rm M_{pl}}$, the model predicts an enhancement in $y$-SD compared to the $\nicefrac{2}{3}$-power-law, $y/y_{\nicefrac{2}{3}} \gtrsim 1$. For $0.1 \lesssim \beta \lesssim 0.23$ and $\phi_{\rm step} \gtrsim 7.54 \, {\rm M_{pl}}$, the $y$-SD values exceed the observational threshold of PIXIE, suggesting a strong potential for detection in this parameter space.

In our analysis, the $\delta$ parameter consistently influences both $\mu$ and $y$ SD, with its value determining the smoothness of the step. This relationship can be understood as follows: when $\delta$ is larger, the step feature in the potential becomes smoother, leading to smaller spectral distortions. Notably, for $\delta > 0.17\, {\rm M_{pl}}$, both $\mu$ and $y$ SD tend towards the values predicted by the standard $\nicefrac{2}{3}$-power-law model, respectively. In the case of $y$-SD, the region is more extensive since for $\delta > 0.08\, {\rm M_{pl}}$, we have $y/y_{\nicefrac{2}{3}} \approx 1$. Conversely, lower values of $\delta$ result in a more abrupt step, which in turn leads to larger spectral distortions. Within the parameter space of $0.05 \lesssim \delta/ {\rm M_{pl}} \lesssim 0.17$, $\mu$-SD values tend to exceed the standard power-law predictions. However, the extent of this deviation also depends on the value of the $\beta$ parameter. When $\delta \lesssim 0.047 \, {\rm M_{pl}}$, we consistently observe stronger distortions regardless of the specific values of $\beta$ and $\phi_{\rm step}$, this is $\mu/\mu_{\nicefrac{2}{3}} > 1$. Furthermore, a combination of $\delta \lesssim 0.047 \, {\rm M_{pl}}$ and $\phi_{\rm step} > 7.54 \, {\rm M_{pl}}$ results in large SD, $y/y_{\nicefrac{2}{3}} > 1$.

We have identified a specific parameter space defined by $\delta < 0.026 \, {\rm M_{pl}}$, $\beta > 0.1$, and $\phi_{\rm step} \gtrsim 7.53 \, {\rm M_{pl}}$. Within this parameter range, $y$-SD signals that potentially exceed the detection threshold of the PIXIE mission are produced. This parameter region also corresponds to the maximum observed values of $\mu$ and $y$ SD. This highlights the potential of this parameter space to provide valuable insights into the most pronounced SD. However, it is important to consider the limitations of PIXIE detection capabilities. This leads to the identification of a parameter space where certain values may not be detectable, for $\mu$-SD, the non-detectable parameter space occurs when $\delta$ exceeds 0.17 \({\rm M_{pl}}\). In this range, the spectral distortions are presumably too subtle. Similarly, for $y$-SD, the undetectable parameter space is defined by $\phi_\text{step} < 7.54 \, {\rm M_{pl}}$, $\beta < 0.10$, and $\delta > 0.026 \, {\rm M_{pl}}$. Within these limits, the distortions are likely below the detection threshold of PIXIE.

Our analysis of SD and the computation of the contributions of $\mu$ and $y$ SD to the photon intensity spectrum highlight the oscillatory nature of the PPS and its potential implications for spectral distortion. This study enriches our understanding and constrains the step-like inflationary models in light of future observational surveys.

\bibliographystyle{unsrt}
\bibliography{sd_step}

\appendix
\section{Extra figures}
In this section, we show some figures that help to the discussion when the $y$--SD is smaller than the expected value for the $\nicefrac{2}{3}$-power-law.
\begin{figure}[H]
     \centering
     \begin{subfigure}[b]{0.49\textwidth}
     \centering
         \includegraphics[width=\textwidth]{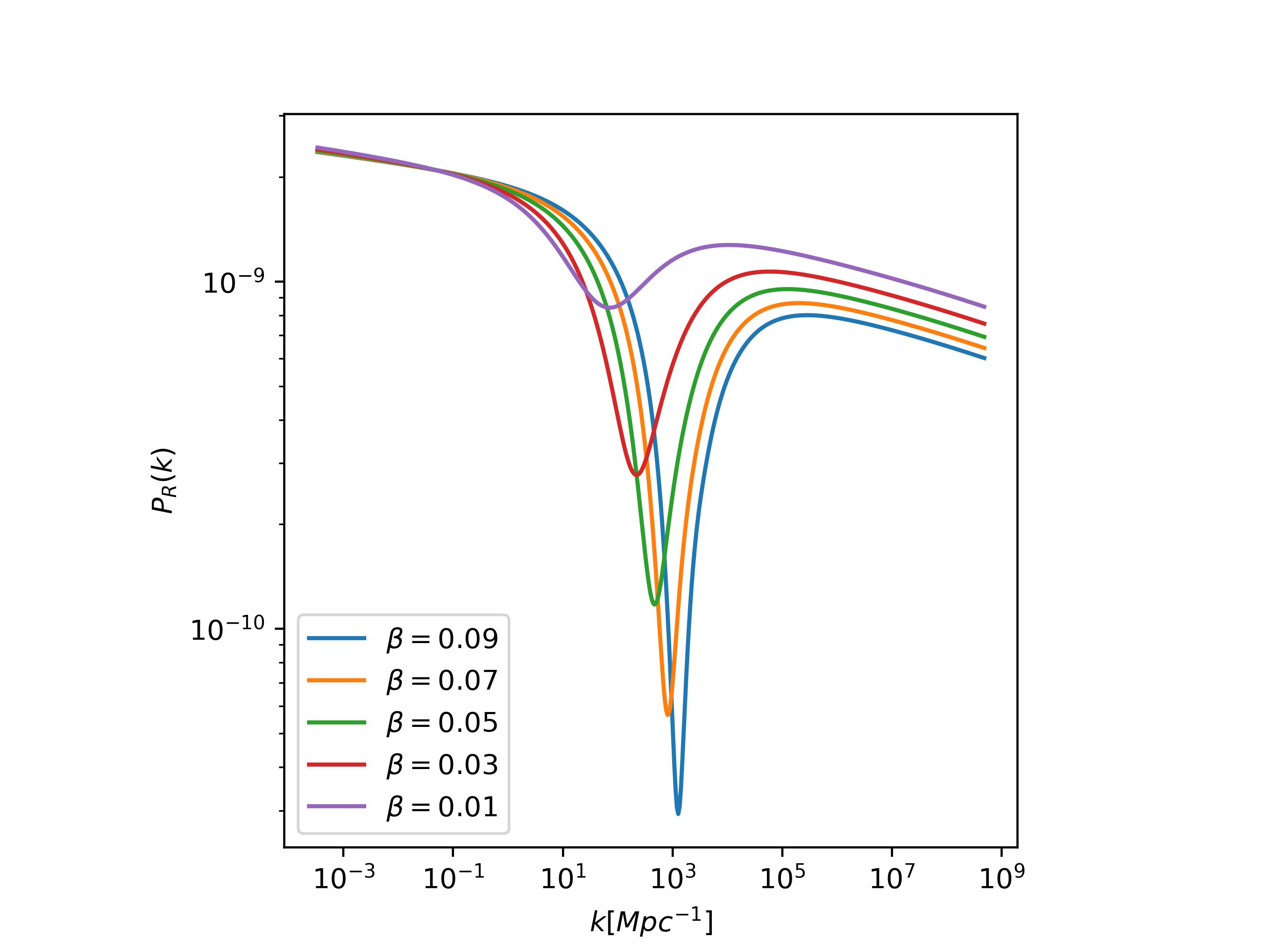}
         \caption{$\phi_{\rm step}$ and $\delta$ fixed }
     \end{subfigure}
     \hfill
     \begin{subfigure}[b]{0.49\textwidth}
     \centering
         \includegraphics[width=\textwidth]{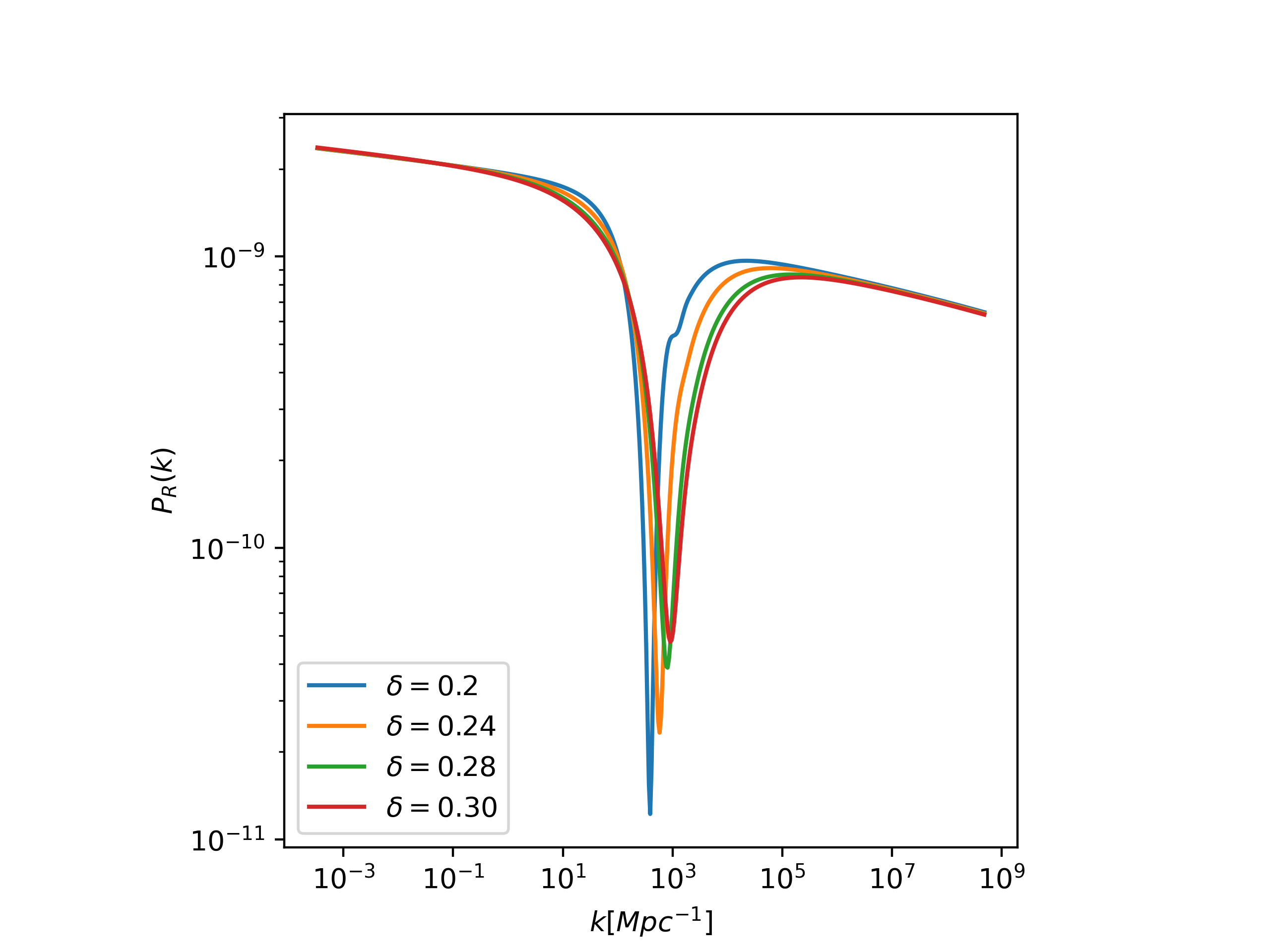}
         \caption{$\phi_\text{step}$ and $\beta$ fixed}
     \end{subfigure}
     \hfill     
     \caption{Scalar primordial power spectrum of inflationary models based in a step potential for small values of $\beta$ (left panel) and for large values of $\delta$ (rigt panel). In both panels we take $\phi_\text{step}=7.55 \, {\rm M_{pl}} $; in the left panel $\delta$ is fixed in $0.3 \, {\rm M_{pl}} $, while, in the right panel $\beta$ is fixed in $0.075$. }
     \label{fig:poso_pps} 
\end{figure}

\end{document}